\documentclass[a4paper,12pt]{article}
\pdfoutput=1
\usepackage{a4wide}
\usepackage[bookmarks=false,colorlinks]{hyperref}
\usepackage[pdftex]{graphicx}
\usepackage{amsmath,amsfonts,amssymb,mathtools,bm}
\usepackage{rotating}
\usepackage{tabularx}
\usepackage{multicol,multirow}
\usepackage{latexsym}
\usepackage{relsize}
\usepackage[center]{subfigure}
\usepackage{float}
\usepackage{caption}
\usepackage{slashed}
\usepackage{lineno}
\usepackage{cite}
\usepackage{appendix}
\usepackage{url}
\usepackage{soul}
\hypersetup{urlcolor=blue, citecolor=blue}

\date{\today}
 
\normalsize

\def\Bbar{\overline{B}}

\def\ubar{\overline{u}}
\def\dbar{\overline{d}}
\def\sbar{\overline{s}}
\def\cbar{\overline{c}}

\def\Qbar{\overline{Q}}

\def\taubar{\overline{\tau}}

\def\nubar{{\overline{\nu}}}
\def\psibar{\overline{\psi}}
\def\Heff{\mathcal{H}_{\rm eff}}
\def\L{\mathcal{L}}

\def\A{\mathcal{A}}
\def\O{\mathcal{O}}
\def\B{\mathcal{B}}
\def\Re{\mathcal{R}e}
\def\Im{\mathcal{I}m}
\def\Dst{{D^*}}
\def\vp{v^\prime}
\def\th{{\theta}}

\allowdisplaybreaks

\begin{document}

\thispagestyle{empty} 
\rightline{KEK-TH-1660}
\rightline{OU-HET 791}
\vspace{2.5cm} 
{\Large
\begin{center}
   {\bf Testing leptoquark models in $\bm{\Bbar \to D^{(*)} \tau\nubar}$}
\end{center}
}
\vspace{0.3cm}

\begin{center}
   {\sc Yasuhito Sakaki, Ryoutaro Watanabe} \\
   \vspace{5mm}
   {\small\emph{Theory Group, IPNS, KEK, Tsukuba, Ibaraki 305-0801, Japan}} \\
   \vspace{5mm}
   {\sc Minoru Tanaka and Andrey Tayduganov} \\
   \vspace{3mm}
   {\small\emph{Department of Physics, Graduate School of Science, Osaka University,}} \\
   {\small\emph{Toyonaka, Osaka 560-0043, Japan}} \\
\end{center}

\vskip3cm
\begin{center}
   \small{\bf Abstract}
\end{center}

\vspace{3mm}
We study potential New Physics effects in the $\Bbar \to D^{(*)} \tau\nubar$ decays. As a particular example of New Physics models we consider the class of leptoquark models and put the constraints on the leptoquark couplings using the recently measured ratios $R(D^{(*)})=\B(\Bbar \to D^{(*)} \tau\nubar)/\B(\Bbar \to D^{(*)} \mu\nubar)$. For consistency, some of the constraints are compared with the ones coming from the current experimental bound on $\B(B \to X_s \nu\nubar)$. In order to discriminate various New Physics scenarios, we examine the correlations between different observables that can be measured in future.


\newpage


\section{Introduction}

Excess of exclusive semitauonic decays of $B$ meson, $\Bbar \to D^{(*)}\tau\nubar$, has been reported by the BaBar and Belle collaborations. In order to test the lepton universality with less theoretical uncertainty, the ratios of the branching fractions are introduced as observables,
\begin{equation}
   R(D^{(*)})\equiv{\B(\Bbar \to D^{(*)}\tau\nubar) \over \B(\Bbar \to D^{(*)}\ell\nubar)} \,,
\end{equation}
where $\ell$ denotes $e$ or $\mu$.
The present experimental results coming from the BaBar experiment are given by \cite{Lees:2012xj,Lees:2013uzd},
\begin{equation}
   R(D)^{\rm BaBar}=0.440\pm 0.072 \,, \quad R(D^*)^{\rm BaBar}=0.332\pm 0.030 \,,
   \label{eq:frombabar}
\end{equation}
with their correlation $\rho=-0.27$, where the statistical and systematical errors are combined assuming Gaussian distribution. For the corresponding results from several Belle publications \cite{Matyja:2007kt,Adachi:2009qg,Bozek:2010xy}, we combine the results which have the smallest errors for each charge mode, and obtain the following numbers:
\begin{equation}
   R(D)^{\rm Belle}=0.390\pm 0.100 \,, \quad R(D^*)^{\rm Belle}=0.347\pm 0.050 \,,
   \label{eq:frombelle}
\end{equation}
where the unknown correlation is assumed to be zero in this case. Further combining Eqs.~\eqref{eq:frombabar} and~\eqref{eq:frombelle}, we obtain 
\begin{equation}
   R(D)=0.421\pm 0.058 \,, \quad R(D^*)=0.337\pm 0.025 \,,
   \label{Eq:combined}
\end{equation}
with the correlation to be $-0.19$. Comparing these experimental results with the Standard Model (SM) predictions,
\begin{equation}
   R(D)^{\rm SM}=0.305\pm 0.012 \,, \quad R(D^*)^{\rm SM}=0.252\pm 0.004 \,,
   \label{Eq:SMP}
\end{equation}
we find that the discrepancy is $3.5\sigma$ combining $R(D)$ and $R(D^*)$.

From the theoretical point of view, two-Higgs-doublet model of type II (2HDM-II) \cite{Gunion:1989we}, which is the Higgs sector of the minimal supersymmetric Standard Model (MSSM) \cite{Martin:1997ns}, has been studied well in the literature \cite{Hou:1992sy,Tanaka:1994ay,Kamenik:2008tj,Nierste:2008qe,Tanaka:2010se} as a candidate of New Physics (NP) that significantly affects the semitauonic $B$ decays. Based on these theoretical works and their experimental data, BaBar collaboration shows that the 2HDM-II is excluded at 99.8\% confidence level (CL)~\cite{Lees:2012xj,Lees:2013uzd}.

This observation has stimulated further theoretical activities for clarifying the origin of the above discrepancy. Several authors have studied various NP scenarios other than 2HDM-II. Possible structures of the relevant four-fermion interaction are identified and models that induce such structures are proposed in the literature \cite{Fajfer:2012vx,Sakaki:2012ft,Datta:2012qk,Bailey:2012jg,Becirevic:2012jf,Fajfer:2012jt,Celis:2012dk,Ko:2012sv,Celis:2013jha,Duraisamy:2013pia}.
One of the interesting four-fermion operators is the scalar type generated in the 2HDMs with flavour changing neutral currents, so-called 2HDM of type III \cite{Crivellin:2012ye}. It is shown that this operator, mentioned as $\O_{S_2}^l$ below, explains the experimental data. Another compelling possibility is the tensor operator $\O_T^l$. Two of us have shown that $\O_T^l$ describes the present experimental results with a reasonable range of its Wilson coefficient and predicts $\tau$ and $D^*$ polarizations different from $\O_{S_2}^l$ \cite{Tanaka:2012nw}. They have also studied a leptoquark model as an intriguing example that induces these operators. The effect of the tensor operator also has been studied recently in Ref. \cite{Biancofiore:2013ki} in a model independent way and in Ref.~\cite{Dorsner:2013tla} in leptoquark models.

In this work, we extend the analysis in Ref.~\cite{Tanaka:2012nw} to all possible leptoquark models \cite{Buchmuller:1986zs}. It is shown that three of them explain the present experimental data quite well. In our study, we carefully investigate theoretical uncertainty in the evaluation of NP contributions in $\Bbar\to D^{(*)}\tau\nubar$ by employing two different sets of relevant hadronic form factors. The rest of the paper is organized as follows: the effective Hamiltonian including all possible four-fermion operators, the relevant helicity amplitudes of $\Bbar \to D^{(*)}\tau\nubar$ and the analytic formulae of differential decay rates are presented in Sec.~\ref{sec:Heff}. After introducing all possible leptoquark models, we evaluate Wilson coefficients of the effective Hamiltonian in Sec.~\ref{sec:LQ}. Constraints from $\Bbar \to D^{(*)}\tau\nubar$ as well as those from $\Bbar \to X_s\nu\nubar$ are also shown in Sec.~\ref{sec:LQ}. Section \ref{sec:LQ} also contains a discussion on theoretical uncertainty in the constraints from $\Bbar \to D^{(*)}\tau\nubar$. In Sec.~\ref{sec:correlations} we study all possible correlations between various observables in order to distinguish different NP models. Section~\ref{sec:conclusions} is devoted to our conclusions. Some details of hadronic form factors and decay distributions are relegated to Appendices.

\section{Effective Hamiltonian and helicity amplitudes}\label{sec:Heff}

Assuming the neutrinos to be left-handed, we introduce the most general effective Hamiltonian that contains all possible four-fermion operators of the lowest dimension for the $b\to c \tau\nubar_l$ transition,
\begin{equation}
   \Heff = {4G_F \over \sqrt2} V_{cb}\left[ (\delta_{l\tau} + C_{V_1}^l)\O_{V_1}^l + C_{V_2}^l\O_{V_2}^l + C_{S_1}^l\O_{S_1}^l + C_{S_2}^l\O_{S_2}^l + C_T^l\O_T^l \right] \,,
      \label{eq:Heff}
\end{equation}
with the operator basis defined as
\begin{equation}
   \begin{split}
      \O_{V_1}^l =& (\cbar_L \gamma^\mu b_L)(\taubar_L \gamma_\mu \nu_{lL}) \,, \\
      \O_{V_2}^l =& (\cbar_R \gamma^\mu b_R)(\taubar_L \gamma_\mu \nu_{lL}) \,, \\
      \O_{S_1}^l =& (\cbar_L b_R)(\taubar_R \nu_{lL}) \,, \\
      \O_{S_2}^l =& (\cbar_R b_L)(\taubar_R \nu_{lL}) \,, \\
      \O_T^l =& (\cbar_R \sigma^{\mu\nu} b_L)(\taubar_R \sigma_{\mu\nu} \nu_{lL}) \,.
   \end{split}
   \label{eq:operators}
\end{equation}
Since the neutrino flavour $l$ is not determined experimentally in $B$ decays, we consider $l=e,\,\mu$ or $\tau$. In the SM, the Wilson coefficients are set to zero, $C_X^l=0$ ($X=V_{1,2},\,S_{1,2},\,T$).

Using this effective Hamiltonian in Eq.~\eqref{eq:Heff} and calculating the  helicity amplitudes (for the details see Ref. \cite{Tanaka:2012nw}), one finds the differential decay rates as follows

\begin{equation}
   \begin{split}
      {d\Gamma(\Bbar \to D \tau\nubar_l) \over dq^2} =& {G_F^2 |V_{cb}|^2 \over 192\pi^3 m_B^3} q^2 \sqrt{\lambda_D(q^2)} \left( 1 - {m_\tau^2 \over q^2} \right)^2 \times\biggl\{ \biggr. \\
                                                      & |\delta_{l\tau} + C_{V_1}^l + C_{V_2}^l|^2 \left[ \left( 1 + {m_\tau^2 \over2q^2} \right) H_{V,0}^{s\,2} + {3 \over 2}{m_\tau^2 \over q^2} \, H_{V,t}^{s\,2} \right] \\
                                                      &+ {3 \over 2} |C_{S_1}^l + C_{S_2}^l|^2 \, H_S^{s\,2} + 8|C_T^l|^2 \left( 1+ {2m_\tau^2 \over q^2} \right) \, H_T^{s\,2} \\
                                                      &+ 3\Re[ ( \delta_{l\tau} + C_{V_1}^l + C_{V_2}^l ) (C_{S_1}^{l*} + C_{S_2}^{l*} ) ] {m_\tau \over \sqrt{q^2}} \, H_S^s H_{V,t}^s \\
                                                      &- 12\Re[ ( \delta_{l\tau} + C_{V_1}^l + C_{V_2}^l ) C_T^{l*} ] {m_\tau \over \sqrt{q^2}} \, H_T^s H_{V,0}^s \biggl.\biggr\} \,,
   \end{split}
\end{equation}
and

\begin{equation}
   \begin{split}
      & {d\Gamma(\Bbar \to \Dst \tau\nubar_l) \over dq^2} = {G_F^2 |V_{cb}|^2 \over 192\pi^3 m_B^3} q^2 \sqrt{\lambda_\Dst(q^2)} \left( 1 - {m_\tau^2 \over q^2} \right)^2 \times\biggl\{ \biggr. \\
      & \quad\quad\quad\quad\quad ( |\delta_{l\tau} + C_{V_1}^l|^2 + |C_{V_2}^l|^2 ) \left[ \left( 1 + {m_\tau^2 \over2q^2} \right) \left( H_{V,+}^2 + H_{V,-}^2 + H_{V,0}^2 \right) + {3 \over 2}{m_\tau^2 \over q^2} \, H_{V,t}^2 \right] \\
      & \quad\quad\quad\quad\quad - 2\Re[(\delta_{l\tau} + C_{V_1}^l) C_{V_2}^{l*}] \left[ \left( 1 + {m_\tau^2 \over 2q^2} \right) \left( H_{V,0}^2 + 2 H_{V,+} H_{V,-} \right) + {3 \over 2}{m_\tau^2 \over q^2} \, H_{V,t}^2 \right] \\
      & \quad\quad\quad\quad\quad + {3 \over 2} |C_{S_1}^l - C_{S_2}^l|^2 \, H_S^2 + 8|C_T^l|^2 \left( 1+ {2m_\tau^2 \over q^2} \right) \left( H_{T,+}^2 + H_{T,-}^2 + H_{T,0}^2  \right) \\
      & \quad\quad\quad\quad\quad + 3\Re[ ( \delta_{l\tau} + C_{V_1}^l - C_{V_2}^l ) (C_{S_1}^{l*} - C_{S_2}^{l*} ) ] {m_\tau \over \sqrt{q^2}} \, H_S H_{V,t} \\
      & \quad\quad\quad\quad\quad - 12\Re[ (\delta_{l\tau} + C_{V_1}^l) C_T^{l*} ] {m_\tau \over \sqrt{q^2}} \left( H_{T,0} H_{V,0} + H_{T,+} H_{V,+} - H_{T,-} H_{V,-} \right) \\
      & \quad\quad\quad\quad\quad + 12\Re[ C_{V_2}^l C_T^{l*} ] {m_\tau \over \sqrt{q^2}} \left( H_{T,0} H_{V,0} + H_{T,+} H_{V,-} - H_{T,-} H_{V,+} \right) \biggl.\biggr\} \,,
   \end{split}
\end{equation}
where $\lambda_{D^{(*)}}(q^2) = ((m_B - m_{D^{(*)}})^2 - q^2)((m_B + m_{D^{(*)}})^2 - q^2)$.

The hadronic amplitudes in $\Bbar \to M \tau \nubar_l$ ($M=D,\,\Dst$) are defined as
\begin{equation}
   \begin{split}
      H_{V_{1,2},\,\lambda}^{\lambda_M}(q^2) =& \varepsilon_\mu^*(\lambda) \, \langle M(\lambda_M) | \cbar \gamma^\mu (1 \mp \gamma_5) b | \Bbar \rangle \,, \\
      H_{S_{1,2},\,\lambda}^{\lambda_M}(q^2) =& \langle M(\lambda_M) | \cbar (1 \pm \gamma_5) b | \Bbar \rangle \,, \\
      H_{T,\,\lambda\lambda^\prime}^{\lambda_M}(q^2) =& -H_{T,\,\lambda^\prime\lambda}^{\lambda_M}(q^2) = \varepsilon_\mu^*(\lambda)\varepsilon_\nu^*(\lambda^\prime) \, \langle M(\lambda_M) | \cbar \sigma^{\mu\nu} (1 - \gamma_5) b | \Bbar \rangle \,, \\
   \end{split}
\end{equation}
where $\lambda_M$ and $\lambda$ denote the meson and virtual intermediate boson helicities ($\lambda_M=s$ and $\lambda_M=0,\pm1$ for $D$ and $\Dst$ respectively, and $\lambda=0,\pm1,t$) in the $B$ rest frame respectively. A detailed description of the matrix elements can be found in Appendix~\ref{app:FF}. The non-zero amplitudes are given below,
\begin{itemize}
   \item $\bm{\Bbar \to D \tau\nubar}$:
\begin{subequations}
   \begin{align}
      \begin{split}
         H_{V,0}^s(q^2) \equiv& \, H_{V_1,0}^s(q^2) = H_{V_2,0}^s(q^2) = \sqrt{\lambda_D(q^2) \over q^2} F_1(q^2) \,, \\
      \end{split} \\
      & \nonumber \\
      \begin{split}
         H_{V,t}^s(q^2) \equiv& \, H_{V_1,t}^s(q^2) = H_{V_2,t}^s(q^2) = {m_B^2-m_D^2 \over \sqrt{q^2}} F_0(q^2) \,, \\
      \end{split} \\
      & \nonumber \\
      \begin{split}\label{eq:HsS_eqm}
         H_S^s(q^2) \equiv& \, H_{S_1}^s(q^2) = H_{S_2}^s(q^2) \simeq {m_B^2-m_D^2 \over m_b-m_c} F_0(q^2) \,, \\
      \end{split} \\
      & \nonumber \\
      \begin{split}
         H_T^s(q^2) \equiv& \, H_{T,+-}^s(q^2) = H_{T,0t}^s(q^2) = -{\sqrt{\lambda_D(q^2)} \over m_B+m_D} F_T(q^2) \,, \\
      \end{split}
   \end{align}
\end{subequations}
\item $\bm{\Bbar \to \Dst \tau\nubar}$:
\begin{subequations}
   \begin{align}
      \begin{split}
         H_{V,\pm}(q^2) \equiv& \, H_{V_1,\pm}^\pm(q^2) = -H_{V_2,\mp}^\mp(q^2) = (m_B+m_\Dst) A_1(q^2) \mp { \sqrt{\lambda_\Dst(q^2)} \over m_B+m_\Dst } V(q^2) \,, \\
      \end{split} \\
      & \nonumber \\
      \begin{split}
         H_{V,0}(q^2) \equiv& \, H_{V_1,0}^0(q^2) = -H_{V_2,0}^0(q^2) = { m_B+m_\Dst \over 2m_\Dst\sqrt{q^2} } \left[ -(m_B^2-m_\Dst^2-q^2) A_1(q^2) \right. \\
                              & \qquad\qquad\qquad\qquad\qquad\qquad\qquad\qquad \left. + { \lambda_\Dst(q^2) \over (m_B+m_\Dst)^2 } A_2(q^2) \right] \,, \\
      \end{split} \\
      & \nonumber \\
      \begin{split}
         H_{V,t}(q^2) \equiv& \, H_{V_1,t}^0(q^2) = -H_{V_2,t}^0(q^2) = -\sqrt{ \lambda_\Dst(q^2) \over q^2 } A_0(q^2) \,, \\
      \end{split} \\
      & \nonumber \\
      \begin{split}\label{eq:HS_eqm}
      H_S(q^2) \equiv& \, H_{S_1}^0(q^2) = -H_{S_2}^0(q^2) \simeq -{ \sqrt{\lambda_\Dst(q^2)} \over m_b+m_c } A_0(q^2) \,, \\
      \end{split} \\
      & \nonumber \\
      \begin{split}
         H_{T,\pm}(q^2) \equiv& \, \pm H_{T,\pm t}^\pm(q^2) = { 1 \over \sqrt{q^2} } \left[ \pm (m_B^2-m_\Dst^2) T_2(q^2) + \sqrt{\lambda_\Dst(q^2)} T_1(q^2) \right] \,, \\
      \end{split} \\
      & \nonumber \\
      \begin{split}
         H_{T,0}(q^2) \equiv& \, H_{T,+-}^0(q^2) = H_{T,0t}^0(q^2) = { 1 \over 2m_\Dst } \left[ -(m_B^2+3m_\Dst^2-q^2) T_2(q^2) \right. \\
                                & \qquad\qquad\qquad\qquad\qquad\qquad\quad\quad\quad \left. + { \lambda_\Dst(q^2) \over m_B^2-m_\Dst^2 } T_3(q^2) \right] \,. \\
      \end{split}
   \end{align}
\end{subequations}
\end{itemize}

In Eqs.~\eqref{eq:HsS_eqm} and \eqref{eq:HS_eqm}, the equations of motion are used for the quark fields.

Up to now all experimental and phenomenological analyses of $\Bbar \to D^{(*)} \tau\nubar$ decays have been made highly relying on the heavy quark effective theory (HQET). Although it provides an extremely useful tool in describing the non-perturbative dynamics of QCD, an alternative description of these decays that does not rely on HQET is welcome. Therefore, in order to be conservative and to estimate the sensitivity of NP constraints to the $\Bbar \to D^{(*)}$ transition matrix elements, two different sets of hadronic form factors are examined:
\begin{itemize}
   \item HQET form factors, parametrized by Caprini {\it et al.} \cite{Caprini:1997mu} with the use of parameters extracted from experiments by the BaBar and Belle collaborations;
   \item form factors, computed by Melikhov and Stech (MS) using relativistic dispersion approach based on the constituent quark model \cite{Melikhov:2000yu}.
\end{itemize}

\section{Testing leptoquark models}\label{sec:LQ}

\subsection{Effective Lagrangian and Wilson coefficients}

Many extensions of the SM, motivated by a unified description of quarks and leptons, predict the existence of new scalar and vector bosons, called leptoquarks, which decay into a quark and a lepton (with model-dependent branching fraction). These particles carry nonzero baryon and lepton numbers, colour and fractional electric charge.

Although for the leptoquark masses that are within experimental reach at collider experiments, the flavour-changing neutral current (FCNC) processes favour leptoquarks that couple to quarks and leptons of the same generation, in this work we study the leptoquarks which couple to the third and the second generation. We use the Lagrangian with the general dimensionless $SU(3)_c\times SU(2)_L\times U(1)_Y$ invariant {\it flavour non-diagonal} couplings of scalar and vector leptoquarks satisfying baryon and lepton number conservation, introduced by Buchm\"{u}ller~{\it et~al.}~\cite{Buchmuller:1986zs}. The interaction Lagrangian that induces contributions to the $b \to c \ell\nubar$ process is given as follows,
\begin{equation}
   \begin{split}
      \L^{\rm LQ} =& \L_{F=0}^{\rm LQ} + \L_{F=-2}^{\rm LQ} \,, \\
                   & \\
      \L_{F=0}^{\rm LQ} =& \left( h_{1L}^{ij}\,\Qbar_{iL} \gamma^\mu L_{jL} + h_{1R}^{ij}\,\dbar_{iR} \gamma^\mu \ell_{jR} \right)U_{1\mu} + h_{3L}^{ij}\,\Qbar_{iL}{\bm\sigma}\gamma^\mu L_{jL}{\bm U}_{3\mu} \\
      &+ \left( h_{2L}^{ij}\,\ubar_{iR} L_{jL} + h_{2R}^{ij}\,\Qbar_{iL} i\sigma_2 \ell_{jR} \right)R_2 \,, \\
      & \\
      \L_{F=-2}^{\rm LQ} =& \left( g_{1L}^{ij}\,\Qbar_{iL}^c i\sigma_2 L_{jL} + g_{1R}^{ij}\,\ubar_{iR}^c \ell_{jR} \right)S_1 + g_{3L}^{ij}\,\Qbar_{iL}^c i\sigma_2{\bm\sigma} L_{jL}{\bm S}_3 \\
                         &+ \left( g_{2L}^{ij}\,\dbar_{iR}^c \gamma^\mu L_{jL} + g_{2R}^{ij}\,\Qbar_{iL}^c \gamma^\mu \ell_{jR} \right)V_{2\mu} \,,
   \end{split}
   \label{eq:LQ_Lagrangian}
\end{equation}
where $Q_i$ and $L_j$ are the left-handed quark and lepton $SU(2)_L$ doublets respectively, while $u_{iR}$, $d_{iR}$ and $\ell_{jR}$ are the right-handed up, down quark and charged lepton $SU(2)_L$ singlets; indices $i$ and $j$ denote the generations of quarks and leptons; $\psi^c = C\psibar^T=C\gamma^0\psi^*$ is a charge-conjugated fermion field. For simplicity, the colour indices are suppressed. The quantum numbers of the leptoquarks are summarized in Table~\ref{tab:LQ_numbers}.

\begin{table}[t]
   \begin{center}
   \begin{tabular}{|c|c|c|c|c|c|c|}
      \hline
                        & $S_1$  & $S_3$  & $V_2$  & $R_2$  & $U_1$  & $U_3$ \\
      \hline
      spin              & 0      & 0      & 1      & 0      & 1      & 1 \\
      \hline
      $F=3B+L$          & -2     & -2     & -2     & 0      & 0      & 0 \\
      \hline
      $SU(3)_c$         & 3$^*$  & 3$^*$  & 3$^*$  & 3      & 3      & 3 \\
      \hline
      $SU(2)_L$         & 1      & 3      & 2      & 2      & 1      & 3 \\
      \hline
      $U(1)_{Y=Q-T_3}$  & 1/3    & 1/3    & 5/6    & 7/6    & 2/3    & 2/3 \\
      \hline
   \end{tabular}
   \caption{\footnotesize Quantum numbers of scalar and vector leptoquarks with $SU(3)_c\times SU(2)_L\times U(1)_Y$ invariant couplings.}
   \label{tab:LQ_numbers}
\end{center}
\end{table}

We note that {\it the fermion fields in Eq. \eqref{eq:LQ_Lagrangian} are given in the gauge eigenstate basis in which Yukawa couplings of the up-type quarks and the charged leptons are diagonal}. Rotating the down-type quark fields into the mass eigenstate basis and performing the Fierz transformations, one finds the general Wilson coefficients {\it at the leptoquark mass scale} for all possible types of leptoquarks contributing to the $b \to c \tau \nubar_l$ process:
\begin{subequations}
   \label{eq:LQ_coefficients}
   \begin{align}
      \label{eq:CV1}
      C_{V_1}^l =& { 1 \over 2\sqrt2 G_F V_{cb} } \sum_{k=1}^3 V_{k3} \left[ {g_{1L}^{kl}g_{1L}^{23*} \over 2M_{S_1^{1/3}}^2} - {g_{3L}^{kl}g_{3L}^{23*} \over 2M_{S_3^{1/3}}^2} + {h_{1L}^{2l}h_{1L}^{k3*} \over M_{U_1^{2/3}}^2} - {h_{3L}^{2l}h_{3L}^{k3*} \over M_{U_3^{2/3}}^2} \right] \,, \\
      C_{V_2}^l =& 0 \,, \\
      \label{eq:CS1}
      C_{S_1}^l =& { 1 \over 2\sqrt2 G_F V_{cb} } \sum_{k=1}^3 V_{k3} \left[ -{2g_{2L}^{kl}g_{2R}^{23*} \over M_{V_2^{1/3}}^2} - {2h_{1L}^{2l}h_{1R}^{k3*} \over M_{U_1^{2/3}}^2} \right] \,, \\
      \label{eq:CS2}
      C_{S_2}^l =& { 1 \over 2\sqrt2 G_F V_{cb} } \sum_{k=1}^3 V_{k3} \left[ -{g_{1L}^{kl}g_{1R}^{23*} \over 2M_{S_1^{1/3}}^2} - {h_{2L}^{2l}h_{2R}^{k3*} \over 2M_{R_2^{2/3}}^2} \right] \,, \\
      \label{eq:CT}
      C_T^l =& { 1 \over 2\sqrt2 G_F V_{cb} } \sum_{k=1}^3 V_{k3} \left[ {g_{1L}^{kl}g_{1R}^{23*} \over 8M_{S_1^{1/3}}^2} - {h_{2L}^{2l}h_{2R}^{k3*} \over 8M_{R_2^{2/3}}^2} \right] \,,
   \end{align}
\end{subequations}
where $V_{k3}$ denotes the CKM matrix elements and the upper index of the leptoquark denotes its electric charge. In the following we will neglect double Cabibbo suppressed $\O(\lambda^2)$ terms and keep only the leading terms proportional to $V_{33}\equiv V_{tb}$.

The vector and axial vector currents are not renormalized and their anomalous dimensions vanish. The scale dependence of the scalar and tensor currents at leading logarithm approximation is given by
\begin{equation}
   \begin{split}
      C_S(\mu_b) =& \left[ \alpha_s(m_t) \over \alpha_s(\mu_b) \right]^{\gamma_S \over 2\beta_0^{(5)}} \left[ \alpha_s(m_{\rm LQ}) \over \alpha_s(m_t) \right]^{\gamma_S \over 2\beta_0^{(6)}} C_S(m_{\rm LQ}) \,, \\ 
      C_T(\mu_b) =& \left[ \alpha_s(m_t) \over \alpha_s(\mu_b) \right]^{\gamma_T \over 2\beta_0^{(5)}} \left[ \alpha_s(m_{\rm LQ}) \over \alpha_s(m_t) \right]^{\gamma_T \over 2\beta_0^{(6)}} C_T(m_{\rm LQ}) \,,
   \end{split}
   \label{eq:QCDrunning}
\end{equation}
where the anomalous dimensions of the scalar and tensor operators are $\gamma_S=-6C_F=-8$, $\gamma_T=2C_F=8/3$ respectively and $\beta_0^{(f)}=11-2n_f/3$ \cite{Dorsner:2013tla}. Taking into account the most recent constraints on the scalar and vector leptoquark masses by the ATLAS and CMS collaborations \cite{ATLAS:2013oea,Chatrchyan:2012sv}, in our numerical analysis we assume that all scalar and vector leptoquarks are of the same mass $m_{\rm LQ}=1$~TeV. The  $b$-quark scale is chosen to be $\mu_b=\overline m_b=4.2$~GeV.

One can easily notice from Eq.~\eqref{eq:LQ_coefficients} that in the simplified scenario with presence of only one type leptoquark, namely $R_2^{2/3}$ or $S_1^{1/3}$, the scalar $C_{S_2}^l$ and tensor $C_T^l$ Wilson coefficients are no longer independent: one finds that at the scale of leptoquark mass $C_{S_2}^l(m_{\rm LQ})=\pm4 C_T^l(m_{\rm LQ})$. Then, using Eq.~\eqref{eq:QCDrunning}, one obtains the relation at the bottom mass scale,
\begin{equation}
   C_{S_2}^l (\overline m_b) \simeq \pm7.8 \, C_T^l (\overline m_b) \,.
\end{equation}

\subsection{Constraints from $\bm{\Bbar\to X_s\nu\nubar}$}

Recent progress in experiment and theory has made FCNCs in $B$ decays good tests of the SM and powerful probes of NP beyond the SM. Along with the $b \to s \gamma$ and $b \to s \ell^+\ell^-$ processes, the $b \to s \nu\nubar$ decay is also sensitive to extensions of the SM. From theoretical point of view, the inclusive decay $\Bbar \to X_s \nu\nubar$ is a very clean process since both perturbative $\alpha_s$ and non-perturbative $1/m_b^2$ corrections are known to be small, what makes it to be well suited to search for NP.

The $b\to s\nu_j\nubar_i$ process can be described by the following effective Hamiltonian,
\begin{equation}
   \Heff = {4G_F \over \sqrt2} V_{tb} V_{ts}^* \left[ \left(\delta_{ij}C_L^{(\rm SM)} + C_L^{ij}\right)\O_L^{ij} + C_R^{ij}\O_R^{ij} \right] \,,
\end{equation}
where the left- and right-handed operators are defined as
\begin{equation}
   \begin{split}
      \O_L^{ij} =& (\sbar_L \gamma^\mu b_L)(\nubar_{jL} \gamma_\mu \nu_{iL}) \,, \\
      \O_R^{ij} =& (\sbar_R \gamma^\mu b_R)(\nubar_{jL} \gamma_\mu \nu_{iL}) \,.
   \end{split}
\end{equation}
In the SM, the Wilson coefficient is determined by box and $Z$-penguin loop diagrams computation which gives,
\begin{equation}
   C_L^{(\rm SM)} = {\alpha \over 2\pi\sin^2\theta_W}X(m_t^2/M_W^2) \,, 
\end{equation}
where the loop function $X(x_t)$ can be found e.g. in Ref.~\cite{Buras:1998raa}.

As one can notice from Eq.~\eqref{eq:LQ_Lagrangian}, the scalar leptoquarks $S_{1,3}^{1/3}$ and vector leptoquarks $V_2^{1/3}$ and $U_3^{-1/3}$ give the following contribution to $b\to s\nu_j\nubar_i$:
\begin{subequations}
   \label{eq:LQ_coeff_Xsnunu}
   \begin{align}
      C_R^{ij} =& -{1 \over 2\sqrt2 G_F V_{tb} V_{ts}^* } \sum_{m,n=1}^3 V_{m3}V_{n2}^* {g_{2L}^{mi}g_{2L}^{nj*} \over M_{V_2^{1/3}}^2} \,, \\
      C_L^{ij} =& -{1 \over 2\sqrt2 G_F V_{tb} V_{ts}^* } \sum_{m,n=1}^3 V_{m3}V_{n2}^* \left[ {g_{1L}^{mi}g_{1L}^{nj*} \over 2M_{S_1^{1/3}}^2} + {g_{3L}^{mi}g_{3L}^{nj*} \over 2M_{S_3^{1/3}}^2} - {2h_{3L}^{ni}h_{3L}^{mj*} \over M_{U_3^{-1/3}}^2} \right] \,.
   \end{align}
\end{subequations}
In the following, for simplicity we neglect the subleading $\O(\lambda)$ terms in Eq. \eqref{eq:LQ_coeff_Xsnunu} and keep only the $V_{tb}V_{cs}^*\simeq1$ term.

One has to note that the $U_3^{-1/3}$ leptoquark does not affect $b\to c\ell\nubar$. In this way, as can be seen from Eq. \eqref{eq:LQ_coefficients}, only the $g_{1(3)L}^{3l}g_{1(3)L}^{23*}$ couplings of the $S_{1(3)}^{1/3}$ leptoquarks can be constrained using both $b \to c \tau\nubar_l$ and $b \to s \nu_\tau\nubar_l$ processes. Nevertheless, assuming that the leptoquarks from the same $SU(2)$ triplet, namely $U_3^{-1/3}$ and $U_3^{2/3}$, have masses of the same order, one can combine the constraints on $h_{3L}^{2l}h_{3L}^{33*}$.

Summing over all neutrino flavours and taking into account that the amplitudes with $i\neq j$ do not interfere with the SM contribution, the branching ratio can be written as
\begin{equation}
   \begin{split}
      {d\B(\Bbar\to X_s\nu\nubar) \over dx} = \tau_B{G_F^2 \over 12\pi^3}|V_{tb} V_{ts}^*|^2 m_b^5 S(x)  \biggl[ 3C_L^{(\rm SM)2} + \sum_{i,j=1}^3\left( |C_L^{ij}|^2 + |C_R^{ij}|^2 \right) \biggr. \\
      \biggl. + 2C_L^{(\rm SM)}\sum_{i=1}^{3}\Re[C_L^{ii*}] \biggr] \,,
   \end{split}
\end{equation}
where $x=E_{\rm miss}/m_b$ and the $S(x)$ function describes the shape of the missing energy spectrum \cite{Grossman:1995gt}. In our estimation we set $m_s=0$ (therefore $1/2\le x \le 1$) and neglect the $\alpha_s$ and $1/m_b^2$ corrections.

\begin{figure}[t!]\centering
   \begin{subfigure}
      {\includegraphics[width=0.3\textwidth]{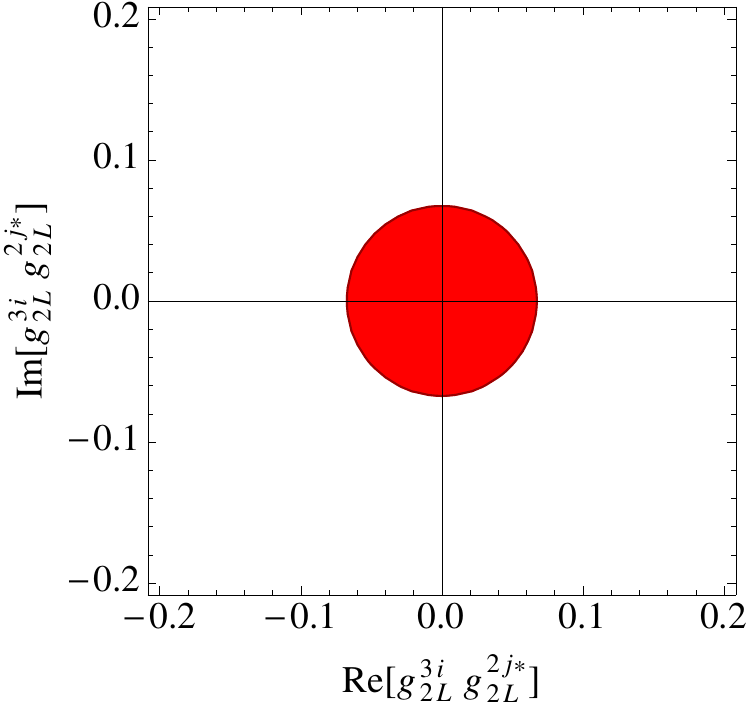}\label{fig:g2Lg2L_Xsnunu}}
      \put(-55,135){\footnotesize(a)}
   \end{subfigure}
      \hspace{3mm}
   \begin{subfigure}
      {\includegraphics[width=0.3\textwidth]{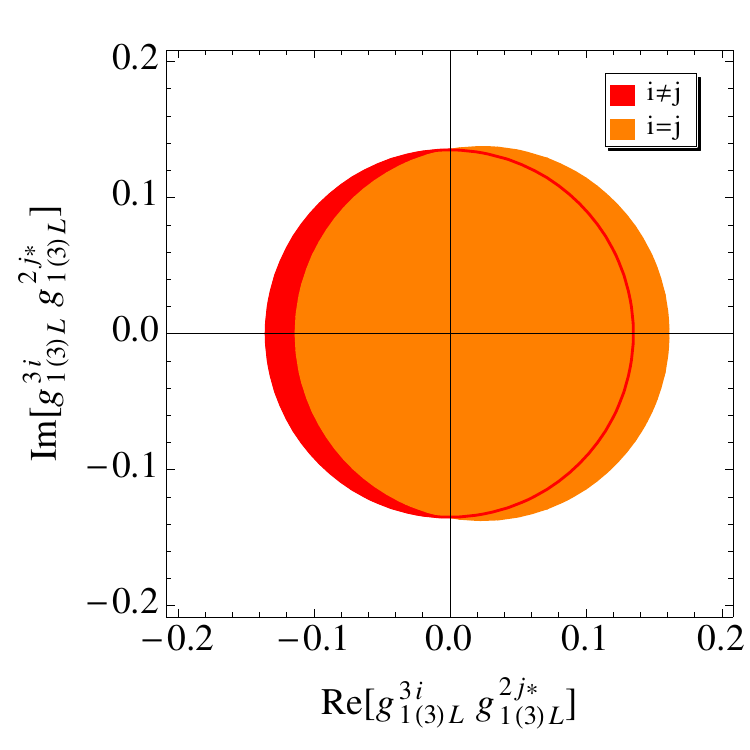}\label{fig:g1Lg1L_Xsnunu}}
      \put(-55,135){\footnotesize(b)}
   \end{subfigure}
      \hspace{3mm}
   \begin{subfigure}
      {\includegraphics[width=0.3\textwidth]{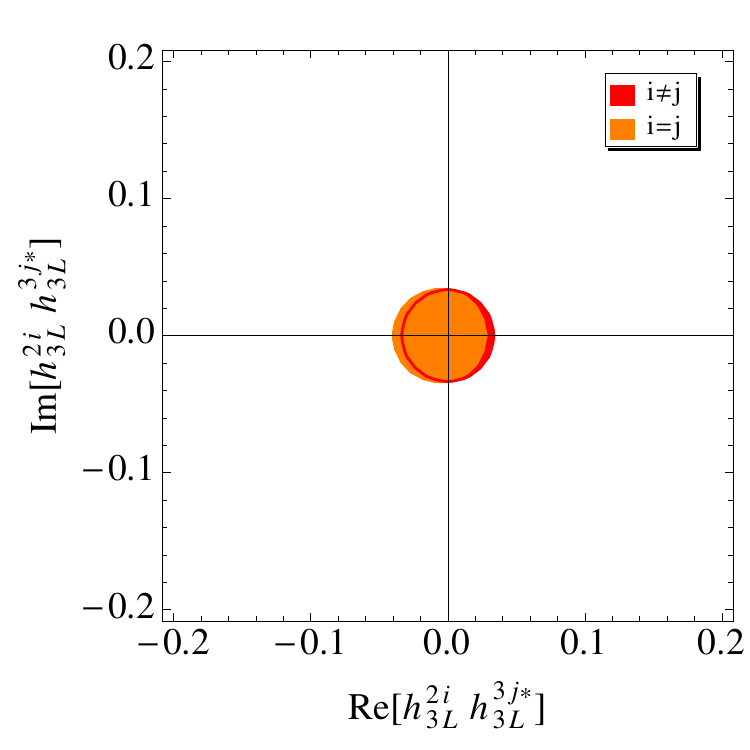}\label{fig:h3Lh3L_Xsnunu}}
      \put(-55,135){\footnotesize(c)}
   \end{subfigure}
   \caption{\footnotesize Constraints on the leptoquark couplings contributing to the $b \to s \nu_j\nubar_i$ process using the experimental upper limit on $\B(B\to X_s\nu\nubar)$}
   \label{fig:LQ_constraints_Xsnunu}
\end{figure}

Using the experimental limit on the inclusive branching ratio, determined by the ALEPH collaboration \cite{Barate:2000rc},
\begin{equation}
   \B^{\rm exp}(B\to X_s\nu\nubar) < 6.4\times10^{-4} \quad \text{at the 90\%~CL} \,,
\end{equation}
and {\it assuming for simplicity that only one specific $ij$ combination of one type of leptoquarks contributes}, we obtain constraints on the leptoquark couplings depicted in Fig.~\ref{fig:LQ_constraints_Xsnunu}. In the case that the couplings are real, the obtained numbers are consistent with the result of Grossman {\it et al.}~\cite{Grossman:1995gt}.

\subsection{Constraints from $\bm{\Bbar \to D \tau\nubar}$ and $\bm{\Bbar \to \Dst \tau\nubar}$}

\begin{figure}[t!]\centering
   \begin{subfigure}
      {\includegraphics[height=0.35\textwidth]{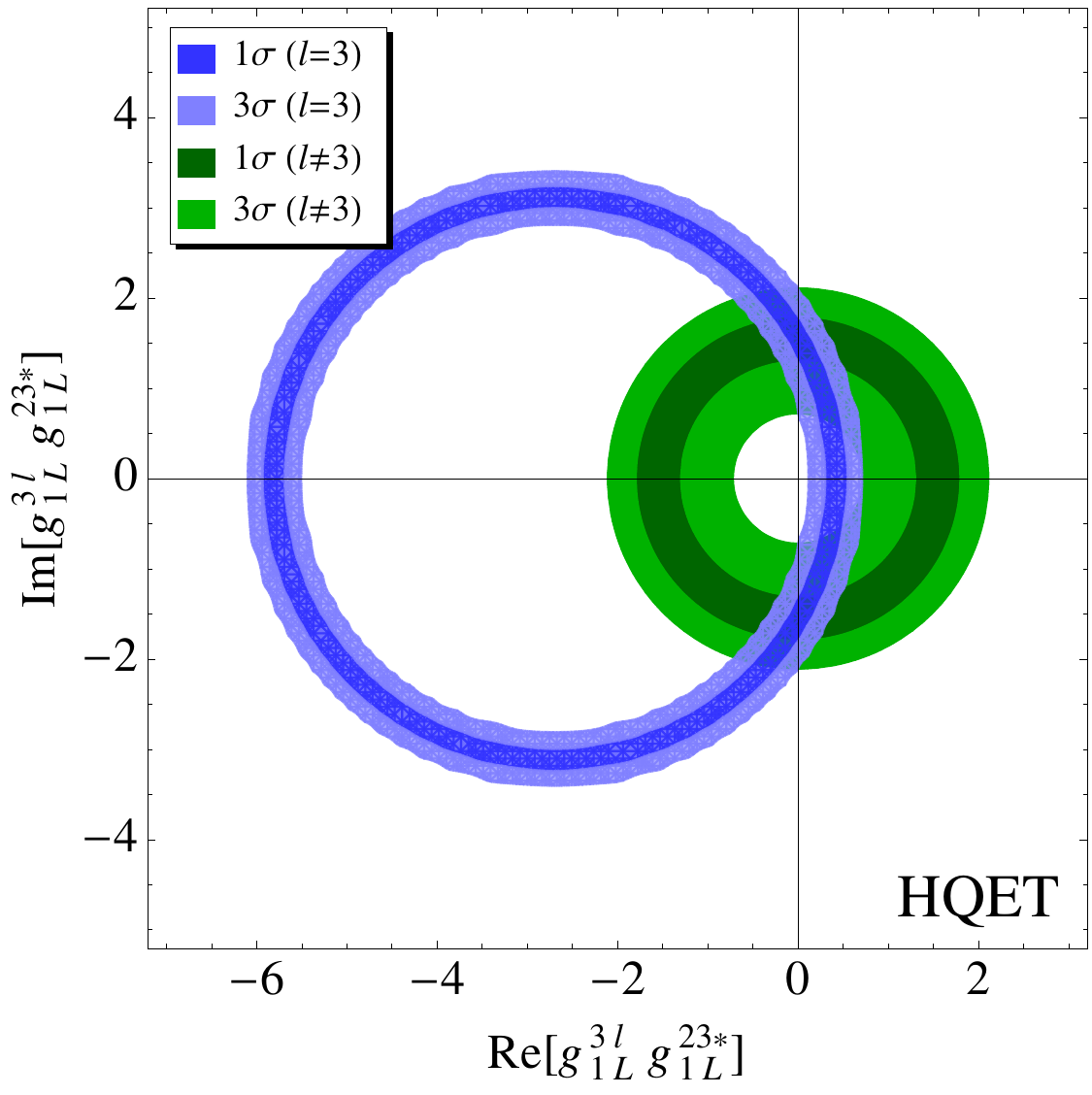}\label{fig:g1Lg1L_HQET}}
      \put(-65,165){\footnotesize(a)}
   \end{subfigure}
   \hspace{5mm}
   \begin{subfigure}
      {\includegraphics[height=0.35\textwidth]{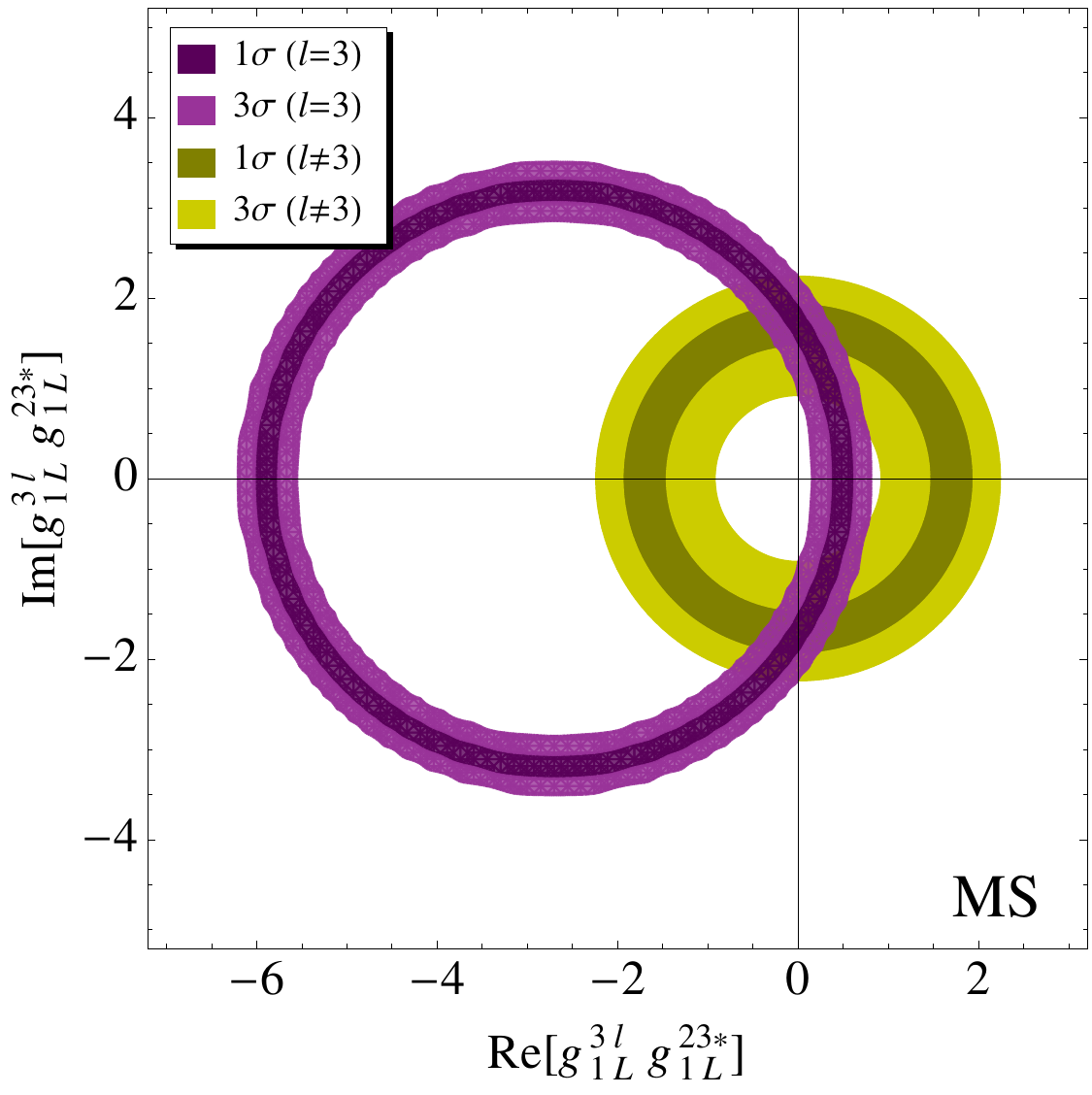}\label{fig:g1Lg1L_MS}}
      \put(-65,165){\footnotesize(b)}
   \end{subfigure}

   \vspace{5mm}
   \begin{subfigure}
      {\includegraphics[height=0.35\textwidth]{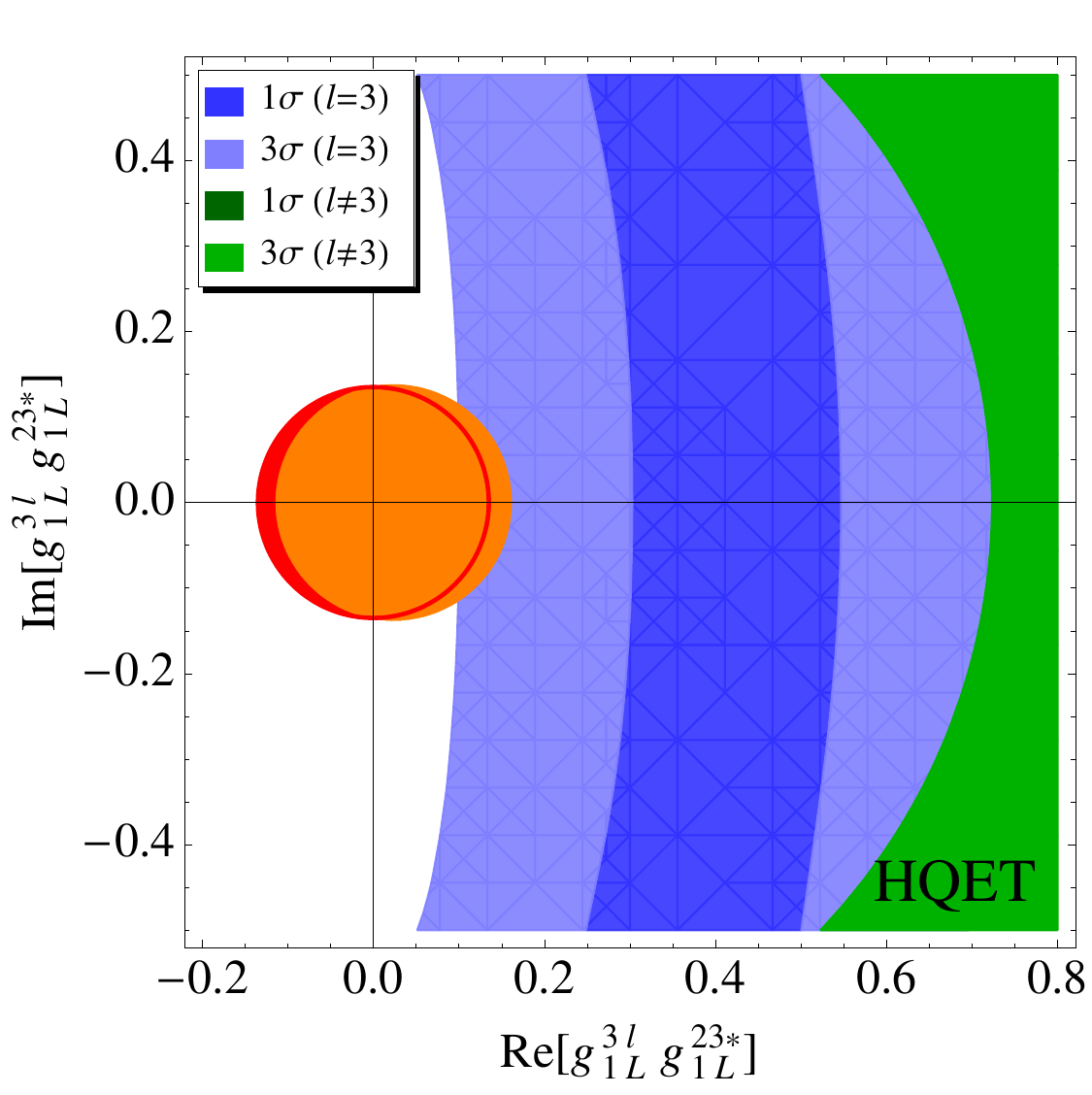}\label{fig:g1Lg1L_zoom_HQET}}
      \put(-65,165){\footnotesize(c)}
   \end{subfigure}
   \hspace{5mm}
   \begin{subfigure}
      {\includegraphics[height=0.35\textwidth]{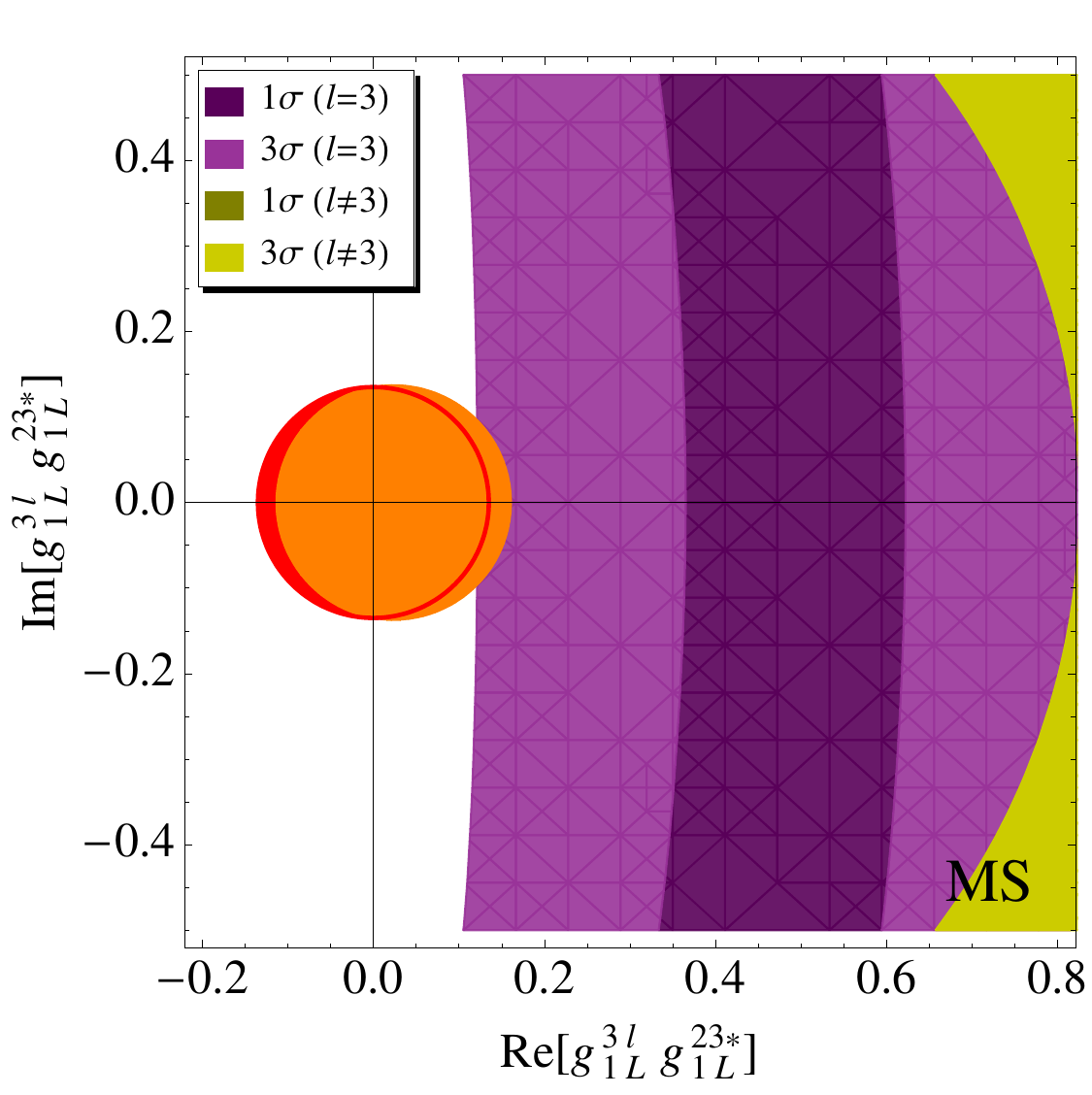}\label{fig:g1Lg1L_zoom_MS}}
      \put(-65,165){\footnotesize(d)}
   \end{subfigure}
   \caption{\footnotesize Constraints on the leptoquark effective couplings at $\mu_b$ scale contributing to the $C_{V_1}$ Wilson coefficient coming from the $\chi^2$ fit of $R(D)$ and $R(\Dst)$. The constraints are obtained by use of form factors evaluated in the HQET (a,c) and the ones computed by Melikhov and Stech (b,d). The zoomed areas around the origin of the plots in (a) and (b) are depicted in (c) and (d) respectively. The orange and red circles show the constraints from the experimental upper limit on $\B(\Bbar \to X_s\nu_\tau\nubar_l)$ for $l=\tau$ and $l\neq\tau$ respectively.}
   \label{fig:LQ_CV1}
\end{figure}

\begin{figure}[t!]\centering
   \begin{subfigure}
      {\includegraphics[height=0.35\textwidth]{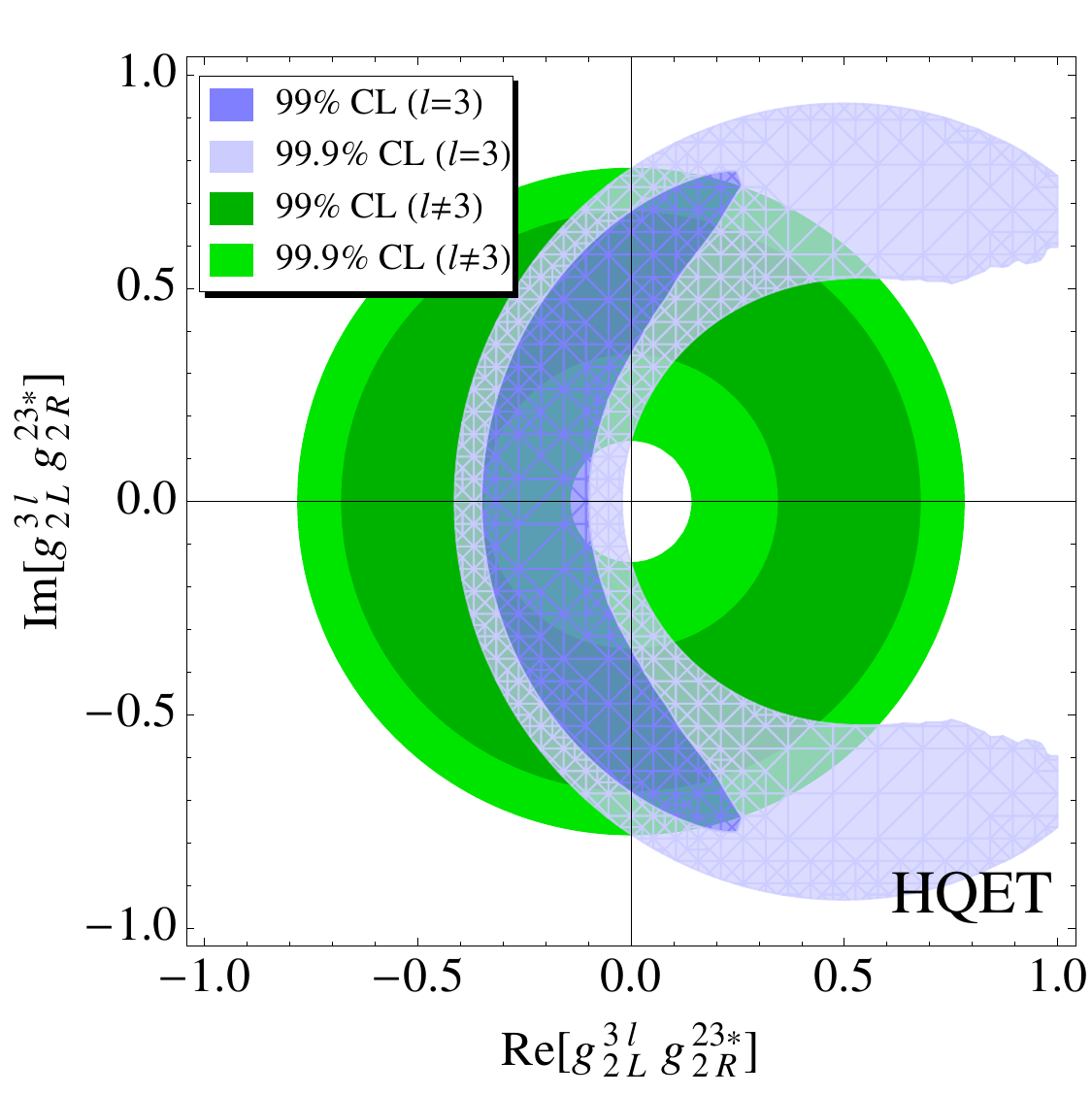}\label{fig:g2Lg2R_HQET}}
      \put(-65,165){\footnotesize(a)}
   \end{subfigure}
   \hspace{5mm}
   \begin{subfigure}
      {\includegraphics[height=0.35\textwidth]{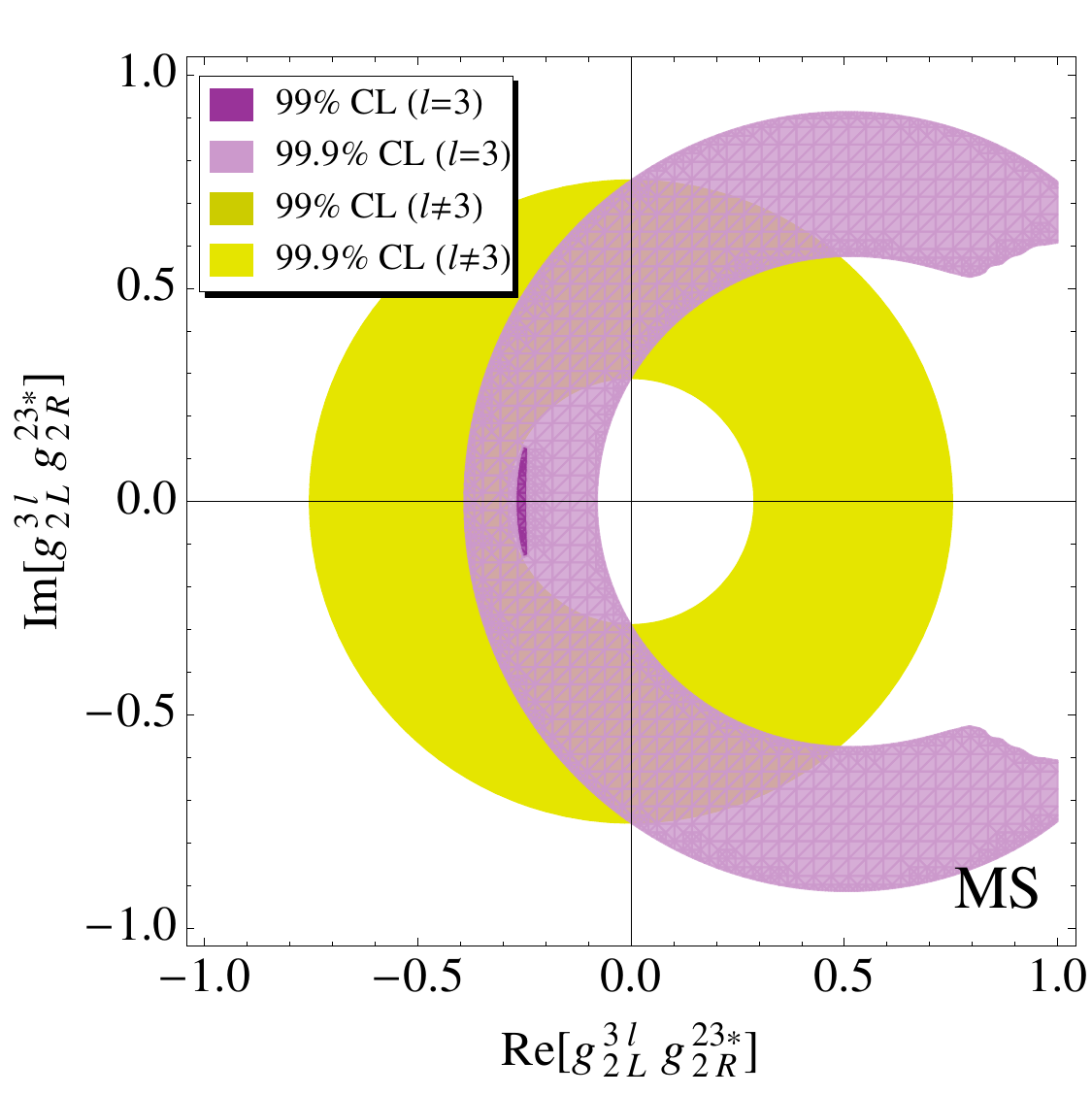}\label{fig:g2Lg2R_MS}}
      \put(-65,165){\footnotesize(b)}
   \end{subfigure}
   \caption{\footnotesize Constraints on the leptoquark effective couplings at $\mu_b$ scale contributing to the $C_{S_1}$ Wilson coefficient coming from the $\chi^2$ fit of $R(D)$ and $R(\Dst)$. The constraints presented in Figs. (a) and (b) are obtained by use of form factors evaluated in the HQET and the ones computed by Melikhov and Stech respectively.}
   \label{fig:LQ_CS1}
\end{figure}

\begin{figure}[t!]\centering
   \begin{subfigure}
      {\includegraphics[height=0.35\textwidth]{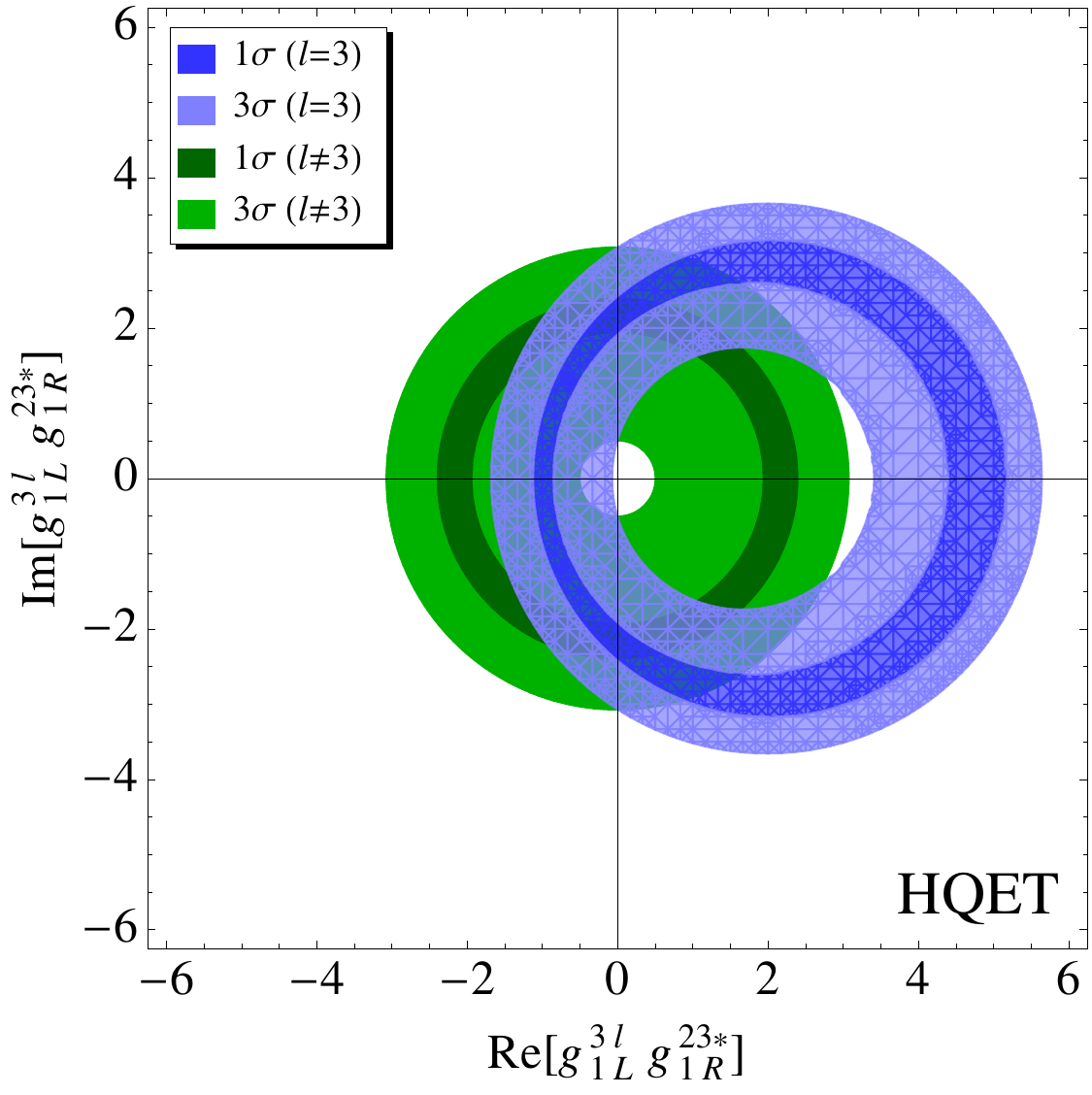}\label{fig:g1Lg1R_HQET}}
      \put(-65,165){\footnotesize(a)}
   \end{subfigure}
   \hspace{5mm}
   \begin{subfigure}
      {\includegraphics[height=0.35\textwidth]{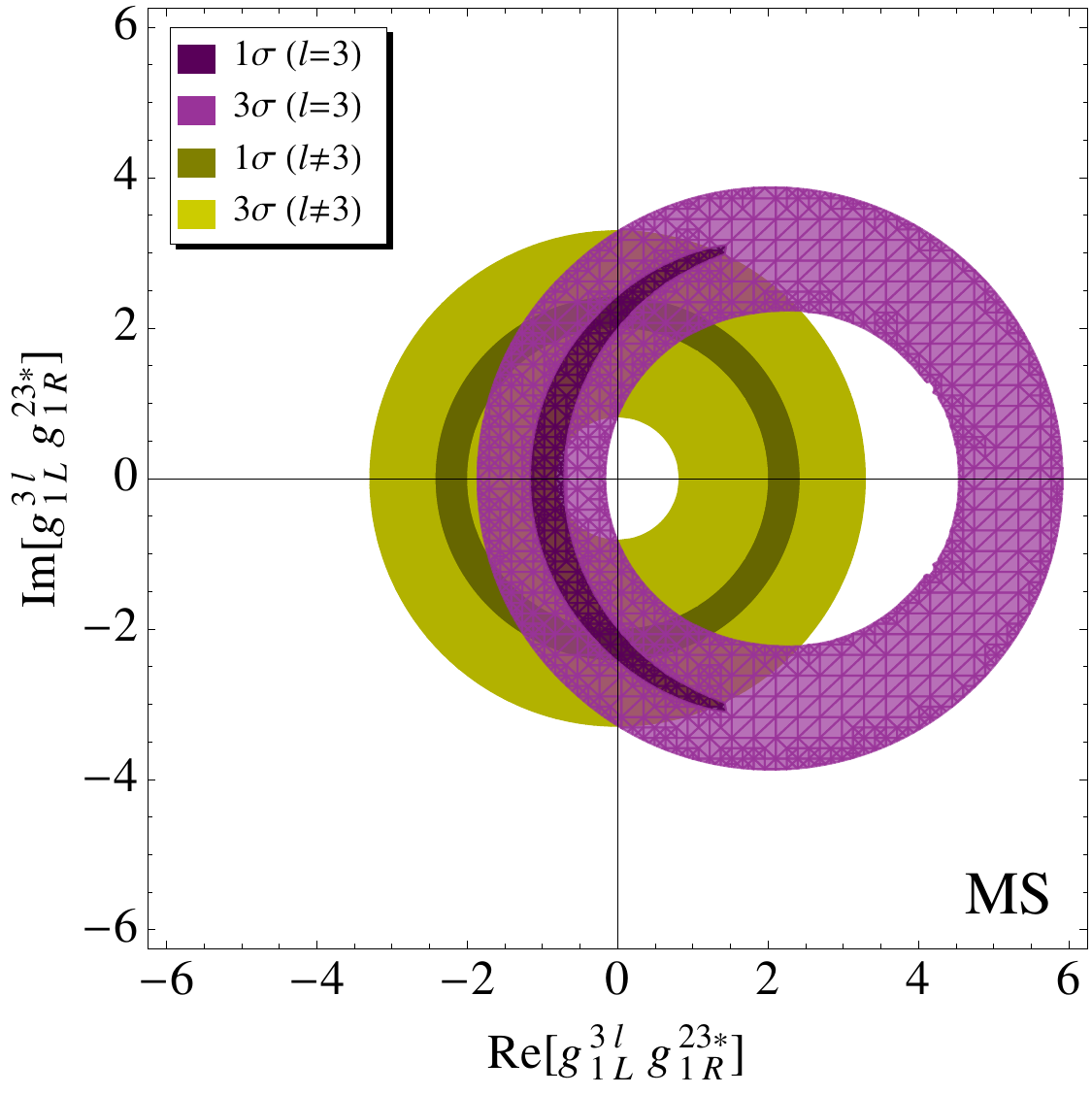}\label{fig:g1Lg1R_MS}}
      \put(-65,165){\footnotesize(b)}
   \end{subfigure}

   \vspace{5mm}
   \begin{subfigure}
      {\includegraphics[height=0.35\textwidth]{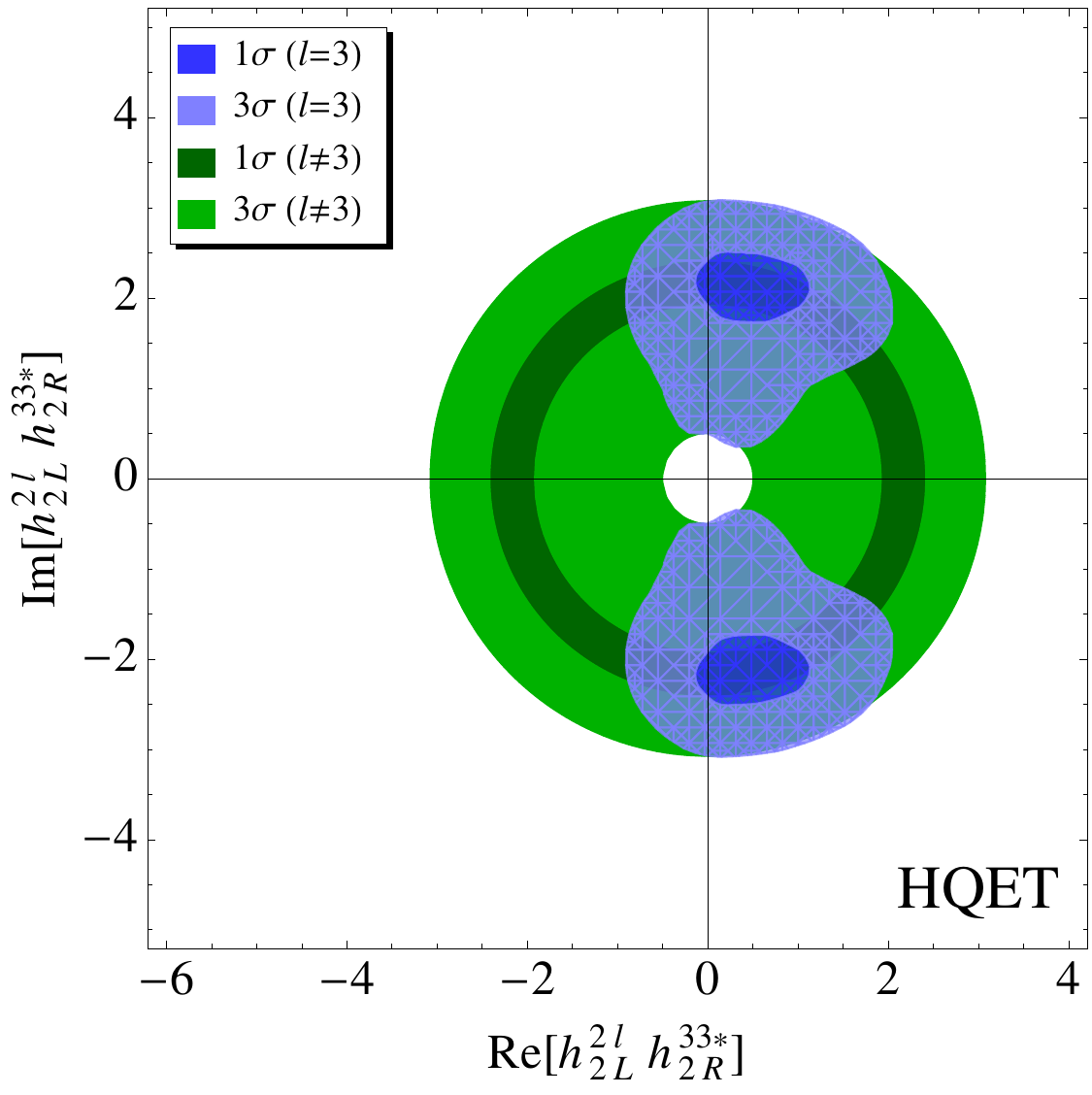}\label{fig:h2Lh2R_HQET}}
      \put(-65,165){\footnotesize(c)}
   \end{subfigure}
   \hspace{5mm}
   \begin{subfigure}
      {\includegraphics[height=0.35\textwidth]{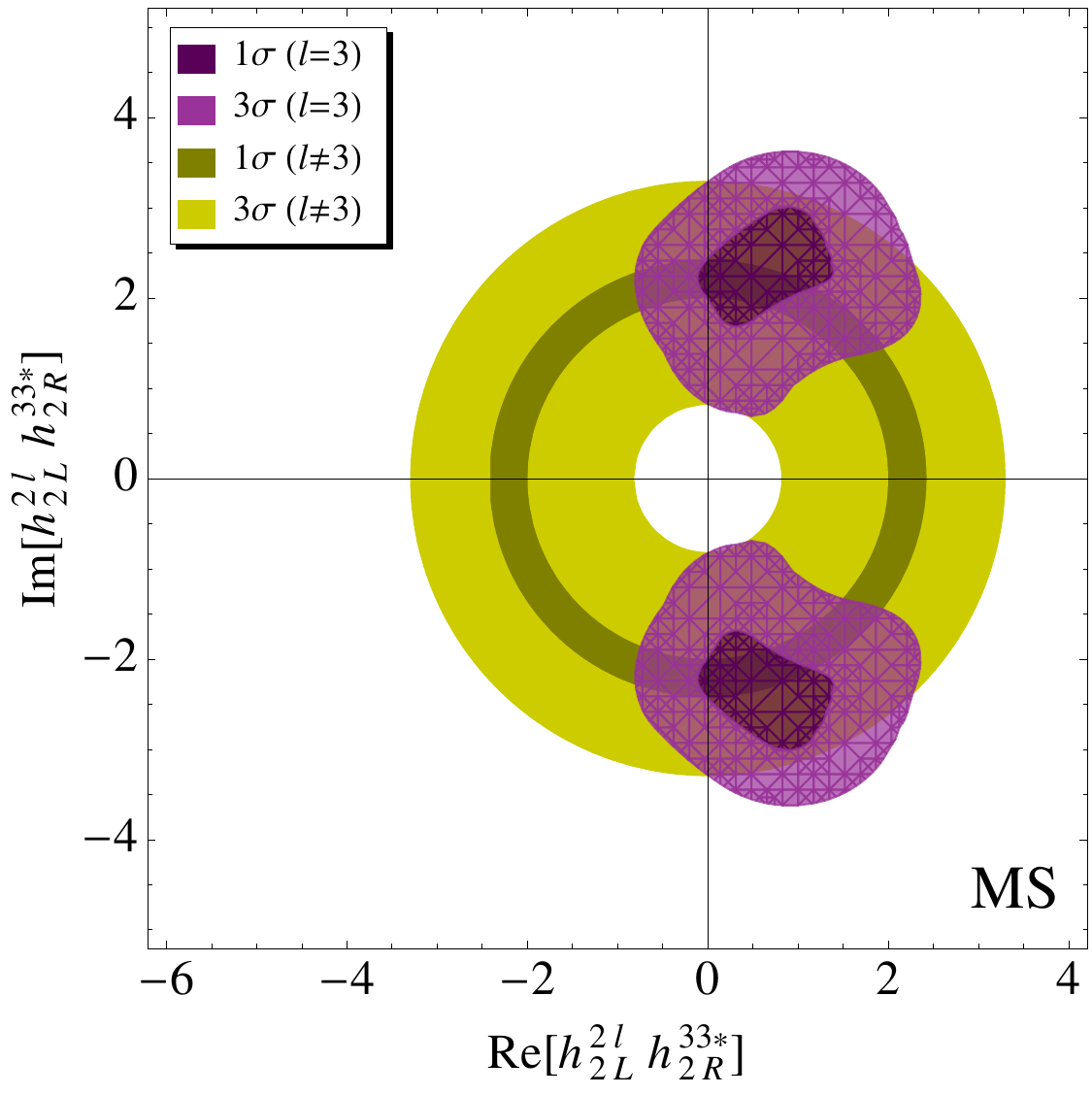}\label{fig:h2Lh2R_MS}}
      \put(-65,165){\footnotesize(d)}
   \end{subfigure}
   \caption{\footnotesize Constraints on the leptoquark effective couplings at $\mu_b$ scale contributing to the $C_{S_2}$ and $C_T$ Wilson coefficients coming from the $\chi^2$ fit of $R(D)$ and $R(\Dst)$. The constraints presented in Figs.~(a,c) and~(b,d) are obtained by use of form factors evaluated in the HQET and the ones computed by Melikhov and Stech respectively.}
   \label{fig:LQ_CS2_CT}
\end{figure}

Using the Wilson coefficients in Eq.~\eqref{eq:LQ_coefficients}, in Figs.~\ref{fig:LQ_CV1}-\ref{fig:LQ_CS2_CT} we provide constraints on various leptoquark effective couplings at the bottom quark mass scale and combine some of them with available bounds coming from $\B(\Bbar \to X_s \nu\nubar)$, discussed in the previous subsection. We consider the general case that the flavour of neutrino is arbitrary. The numerical results of two different sets of form factors are shown for comparison, including the theoretical uncertainties sketched in Appendix~\ref{app:FF}.

In Fig.~\ref{fig:LQ_CV1}, as an example, we present the constraints on the $g_{1L}^{3l}g_{1L}^{23*}$ product of couplings of the $S_1^{1/3}$ leptoquark assuming that the other couplings are zero. The other constraints on $g_{3L}^{3l}g_{3L}^{23*}$ of the $S_3^{1/3}$ leptoquark and $h_{1(3)L}^{2l}h_{1(3)L}^{33*}$ of the $U_{1(3)}^{2/3}$ leptoquark can be easily obtained by rescaling and/or reflecting the constraints from $R(D^{(*)})$ in Figs.~\ref{fig:LQ_CV1} (see Eq.~\eqref{eq:CV1}).

Figures~\ref{fig:g1Lg1L_zoom_HQET} and \ref{fig:g1Lg1L_zoom_MS} represent the zoomed areas around the origin of the plots in Figs.~\ref{fig:g1Lg1L_HQET} and \ref{fig:g1Lg1L_MS} respectively, combined with the constraints from Figs.~\ref{fig:g1Lg1L_Xsnunu}. As one can notice, the case of $\nubar_l \neq \nubar_\tau$ is excluded since the constraints on $g_{1L}^{3l}g_{1L}^{23*}$ coming from $\Bbar \to D^{(*)} \tau\nubar$ and $\Bbar \to X_s \nu\nubar$ are inconsistent, namely there is no overlap between red and green/yellow allowed regions in Figs.~\ref{fig:g1Lg1L_zoom_HQET}/\ref{fig:g1Lg1L_zoom_MS}. The results for the case of $\nubar_l = \nubar_\tau$ are consistent only at $3\sigma$ level and force the couplings to be rather small. For other models, the similar conclusion can be made for $g_{3L}^{3l}g_{3L}^{23*}$ and $h_{3L}^{2l}h_{3L}^{33*}$.
On the contrary, the $U_1^{2/3}$ leptoquark couplings, $h_{1L}^{2l}h_{1L}^{33*}$, remain unconstrained from $\Bbar\to X_s \nu\nubar$ and the magnitude of the order of $O(1)$ can be sufficient to explain the current measurements of $R(D)$ and $R(\Dst)$.

We find that the model with the vector $V_2^{1/3}$ leptoquark exchange with $g_{2L}^{3l}g_{2R}^{23*}$ couplings is hardly possible due to the low compatibility with the experimental data as can be seen from Fig.~\ref{fig:LQ_CS1}. We note that the allowed regions of 99\% CL and 99.9\% CL are shown in Fig.~\ref{fig:LQ_CS1} since there is no allowed region even at 95\% CL. The $h_{1L}^{2l}h_{1R}^{33*}$ couplings of the $U_1^{2/3}$ leptoquark have the same allowed space as $g_{2L}^{3l}g_{2R}^{23*}$ of the $V_2^{1/3}$ leptoquark in Fig.~\ref{fig:LQ_CS1} (see Eq.~\eqref{eq:CS1}).

In Fig.~\ref{fig:LQ_CS2_CT} we demonstrate that the scalar $S_1^{1/3}$ and $R_2^{2/3}$ leptoquark effective couplings, $g_{1L}^{3l}g_{1R}^{23*}$ and $h_{2L}^{2l}h_{2R}^{33*}$, of $O(1)$ are sufficient to explain the present data for the leptoquark mass scale of the order of 1~TeV. It is interesting to note from Figs.~\ref{fig:h2Lh2R_HQET} and \ref{fig:h2Lh2R_MS} that the $R_2^{2/3}$ leptoquark couplings are favoured to be purely imaginary what could be tested directly by studying $\chi$ angular distribution in $\Bbar \to \Dst (\to D\pi) \tau\nubar$ (where $\chi$ is the azimuthal angle between the planes formed by the $W-\tau$ and $\Dst-D$ systems in the $\Bbar$ rest frame).

\subsection{Sensitivity of the constraints to hadronic form factors}\label{sec:FF_discussion}

To conclude this section, we discuss the sensitivity of the NP constraints to hadronic form factors and their theoretical uncertainties. In Figs.~\ref{fig:LQ_CV1}-\ref{fig:LQ_CS2_CT} we show the comparison of the resulting constraints on leptoquark effective couplings, obtained by using the form factors evaluated in the HQET by Caprini {\it et al.} \cite{Caprini:1997mu} and the ones computed by Melikhov and Stech in the constituent quark model \cite{Melikhov:2000yu}.  These two sets have fairly different uncertainties although both of them describe the experimental results of $\Bbar \to D^{(*)}\ell\nubar$ and are consistent with the heavy quark symmetry.

We find that both sets of form factors give similar allowed regions in the parameter space for most of leptoquark models. The constraints on the product of couplings of the scalar $S_{1(3)}^{1/3}$ and vector $U_{1(3)}^{2/3}$ leptoquarks with only left-handed couplings ($g_{1(3)L}^{3l}g_{1(3)L}^{23*}$ and $h_{1(3)L}^{2l}h_{1(3)L}^{33*}$ respectively) in Fig.~\ref{fig:LQ_CV1} look practically identical and therefore the effect of the choice of the form factor set is negligible.

In our study we observe that in the case of the vector $V_2^{1/3}$ and $U_1^{2/3}$ leptoquarks with both left- and right-handed couplings ($g_{2L}^{3l}g_{2R}^{23*}$ and $h_{1L}^{2l}h_{1R}^{33*}$ respectively), the degree of exclusion highly depends on the employed form factors (see Fig.~\ref{fig:LQ_CS1}). One can notice from Fig.~\ref{fig:g2Lg2R_MS} that for the case of the MS form factors there is practically no allowed region at 99\%~CL what makes this model disfavoured. This means that we must be careful about theoretical uncertainties when excluding NP models.

In Fig.~\ref{fig:LQ_CS2_CT} we show the resulting constraints on the scalar $S_1^{1/3}$ and $R_2^{2/3}$ leptoquark effective couplings ($g_{1L}^{3l}g_{1R}^{23*}$ and $h_{2L}^{2l}h_{2R}^{33*}$ respectively) which contribute to both $C_{S_2}$ and $C_T$ Wilson coefficients and therefore are sensitive to tensor form factors. One can notice that, compared to Fig.~\ref{fig:LQ_CV1}, the constraints in Fig.~\ref{fig:LQ_CS2_CT} look slightly different for two sets of form factors. The form factor uncertainty tends to cancel in the ratios $R(D^{(*)})$ for the case of the SM-like operators, $\O_{V_1}^l$, as can be seen in Fig.~\ref{fig:LQ_CV1}. On the other hand, we do not expect such cancellation in the case of the scalar and tensor operators, $\O_{S_{1,2}}^l$ and $\O_T^l$. This  makes the NP constraints to be more sensitive to the tensor form factor uncertainties and hence can explain the difference between the HQET and MS results in Fig.~\ref{fig:LQ_CS2_CT}.

In Table~\ref{tab:Im-Re_numbers} we give explicitly some numerical results for the allowed parameter space compatible with the experimental data at $1\sigma$ level (except for the vector $V_2^{1/3}$ leptoquark couplings $g_{2L}^{33}g_{2R}^{23*}$, for which we present the ranges at 99\%~CL due to the absence of the allowed space at $1\sigma$ and $2\sigma$ levels). For illustration, we assume the product of couplings to be purely real or purely imaginary. As one can see from Table~\ref{tab:Im-Re_numbers}, the allowed ranges are well consistent for two sets of form factors. The exception is the $V_2^{1/3}$ leptoquark couplings $g_{2L}^{33}g_{2R}^{23*}$ which have a very tiny parameter space at 99\% CL for the MS form factors.

Incidentally, we would like to note that the HQET parameters, $\rho_{D,\Dst}^2$ and $R_{1,2}(1)$ (see Appendix \ref{app:HQET}), are extracted from experiments by the BaBar and Belle collaborations~\cite{Aubert:2007rs,Aubert:2008yv,Abe:2001yf,Dungel:2010uk} {\it assuming only the SM contribution to the total amplitude of $\Bbar \to D^{(*)}\ell\nubar_\ell$ ($\ell=e,\mu$)}. Therefore, in order to use the fitted HQET form factors, one has to make an important assumption that couplings of NP particles to light leptons are significantly suppressed as in the 2HDM-II and NP effects can be observed only in the tauonic decay modes.

\begin{table}[t]
   \begin{center}
   \begin{tabular}{|c|c|c|}
      \hline
      & HQET & MS \\
      \hline
      \begin{tabular}{c}
         $|\Im[h_{2L}^{23}h_{2R}^{33*}]|$ \\
         $|\Im[g_{1L}^{33}h_{1R}^{23*}]|$
      \end{tabular} & $[1.92; 2.42]$ & $[1.99; 2.44]$ \\
      \hline
      $\Re[g_{1L}^{33}g_{1R}^{23*}]$ & \begin{tabular}{c}
                                          $[-1.12; -0.85]$ \\
                                          $[4.40; 5.17]$
                                       \end{tabular} &
                                       \begin{tabular}{c}
                                          $[-1.16; -0.71]$
                                       \end{tabular} \\
      \hline
      $\Re[h_{1L}^{23}h_{1L}^{33*}]$ & \begin{tabular}{c}
                                          $[-2.97; -2.85]$ \\
                                          $[0.15; 0.27]$
                                       \end{tabular} &
                                       \begin{tabular}{c}
                                          $[-3.01; -2.88]$ \\
                                          $[0.18; 0.31]$
                                       \end{tabular} \\
      \hline
      $|\Im[h_{1L}^{23}h_{1L}^{33*}]|$ & $[0.65; 0.90]$ & $[0.73; 0.97]$ \\
      \hline
      $\Re[g_{2L}^{33}g_{2R}^{23*}]$ & \begin{tabular}{c}
                                          $[-0.35; -0.10]$
                                       \end{tabular} &
                                       \begin{tabular}{c}
                                          $[-0.27; -0.24]$
                                       \end{tabular} \\
      \hline
      $|\Im[g_{2L}^{33}g_{2R}^{23*}]|$ & $[0.34; 0.68]$ & \\
      \hline
   \end{tabular}
   \caption{\footnotesize Comparison of the $\pm1\sigma$ allowed ranges for the leptoquark effective couplings using the form factors evaluated in the HQET and the ones computed by Melikhov and Stech. The intervals for $g_{2L}^{33}g_{2R}^{23*}$ are given at 99\% CL level due to the absence of the allowed space at $1\sigma$ and $2\sigma$ levels. The products of couplings are assumed to be purely real or imaginary.}
   \label{tab:Im-Re_numbers}
\end{center}
\end{table}

\section{Correlations between observables}\label{sec:correlations}

In order to distinguish various NP models, we study the following observables which could be sensitive to NP:
\begin{itemize}
   \item $\tau$ forward-backward asymmetry,
      \begin{equation}
            \A_{\rm FB} = { \int_0^1 {d\Gamma \over d\cos\th}d\cos\th-\int^0_{-1}{d\Gamma \over d\cos\th}d\cos\th \over \int_{-1}^1 {d\Gamma \over d\cos\th}d\cos\th } = { \int b_\th(q^2) dq^2 \over \Gamma } \,,
         \end{equation}
         where $\theta$ is the angle between the three-momenta of $\tau$ and $\Bbar$ in the $\tau\nubar$ rest frame.

   \item $\tau$ polarization parameter by studying further $\tau$ decays,
      \begin{equation}
         P_\tau = { \Gamma(\lambda_\tau=1/2) - \Gamma(\lambda_\tau=-1/2) \over \Gamma(\lambda_\tau=1/2) + \Gamma(\lambda_\tau=-1/2) } \,,
         \label{eq:Ptau}
         \end{equation}

   \item $\Dst$ longitudinal polarization using the $\Dst\to D\pi$ decay,
      \begin{equation}
         P_\Dst = { \Gamma(\lambda_\Dst=0) \over \Gamma(\lambda_\Dst=0) + \Gamma(\lambda_\Dst=1) + \Gamma(\lambda_\Dst=-1) } \,.
         \label{eq:PDst}
      \end{equation}
\end{itemize}
Here, for shortness, $\Gamma$ denotes $\Gamma(\Bbar \to D^{(*)} \tau \nubar)$. The $q^2$ distributions for various $\tau$ and $\Dst$ polarization states together with $b_\theta(q^2)$ can be found in Appendix~\ref{app:distrib_and_pol}.

In order to determine $\theta$ angle, the $\tau$ momentum reconstruction is necessary. It is not apparent whether this is possible due to the two or more missing neutrinos in the decay modes under consideration \cite{Tanaka:1994ay}. Here we mention a proposal in LHCb to utilize the information on the vertices of $\Bbar$ and $\tau$ production/decay for identifying $\Bbar \to D^{*}\tau\nubar$ process in their environment \cite{John:2012,Keune:2012phd}. The $\tau$ production/decay vertex information, which can be obtained using the $\Dst \to D\pi \,/\, \tau \to 3h \, \nu$ decays, allows us to determine the three-momentum of $\tau$ in the lab frame with a two-fold ambiguity. Then, the same solution can be applied for the $\Bbar$ meson case, knowing the $\Bbar$ production/decay vertices and the $\tau$ momentum. As a result, performing a boost to the $\tau\nubar$ rest frame, $\theta$ can be determined with a four-fold ambiguity. If a similar technique is available at super~$B$ factories, this ambiguity can be reduced to a two-fold one due to the full knowledge of the initial $B$ meson kinematics.

The longitudinal $\tau$ polarization is measurable without reconstructing the $\tau$ momentum as is discussed in Ref.~\cite{Tanaka:2010se}. The expected precision at super~$B$ factories with $50\,\mathrm{ab}^{-1}$ is $\delta P_\tau \sim 0.04(0.03)$ for the $D^{(*)}$ mode. The $D^*$ polarization is also measurable from the pion distribution in the $D^*$ decay. The precision at super~$B$ factories with $50\ \mathrm{ab}^{-1}$ is estimated as $\delta P_{D^*} \sim 5\times10^{-3}$.

\begin{figure}[t!]\centering
      \includegraphics[height=0.31\textwidth]{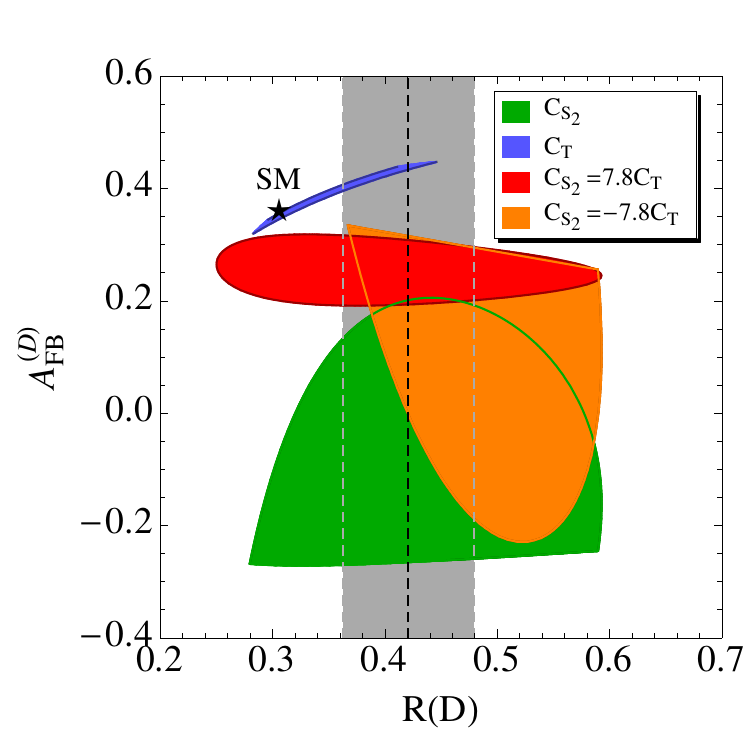} \hspace{2mm}
      \includegraphics[height=0.30\textwidth]{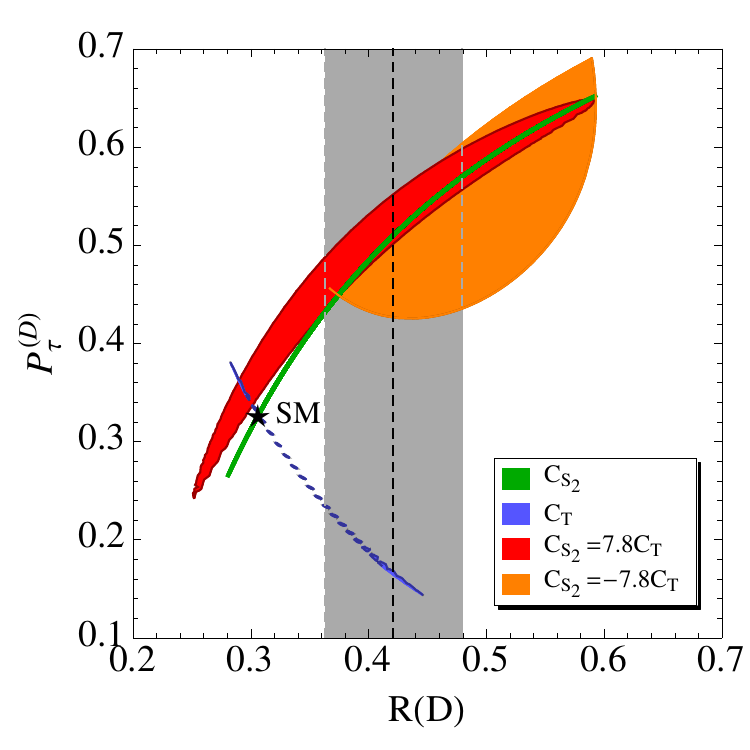} \hspace{2mm}
      \includegraphics[height=0.29\textwidth]{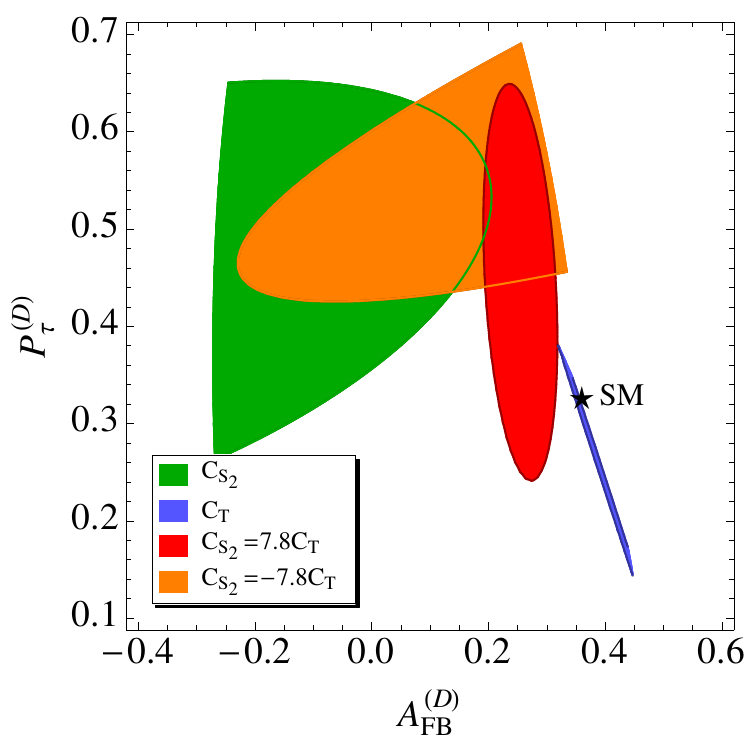}
      \\
      \vspace{2mm}
      \includegraphics[height=0.31\textwidth]{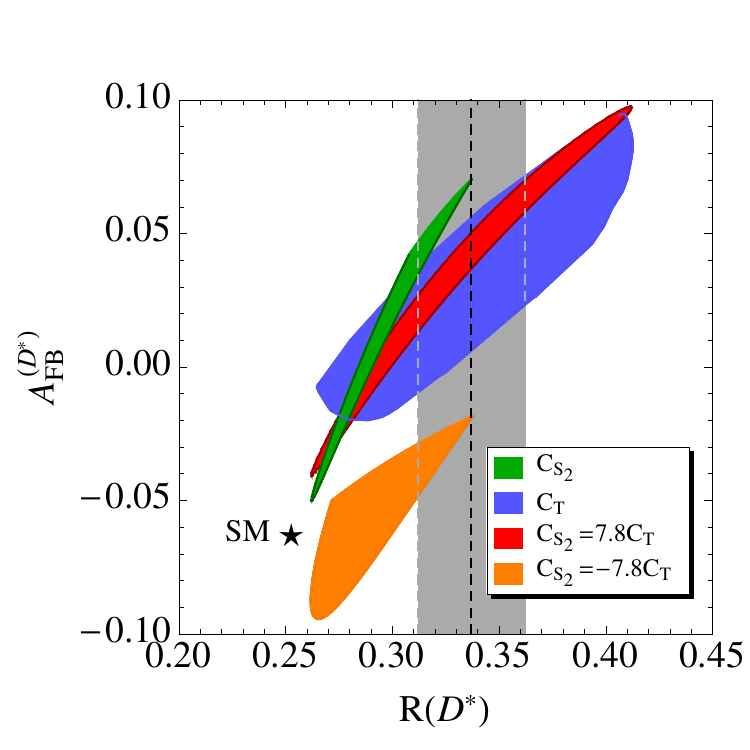} \hspace{2mm}
      \includegraphics[height=0.30\textwidth]{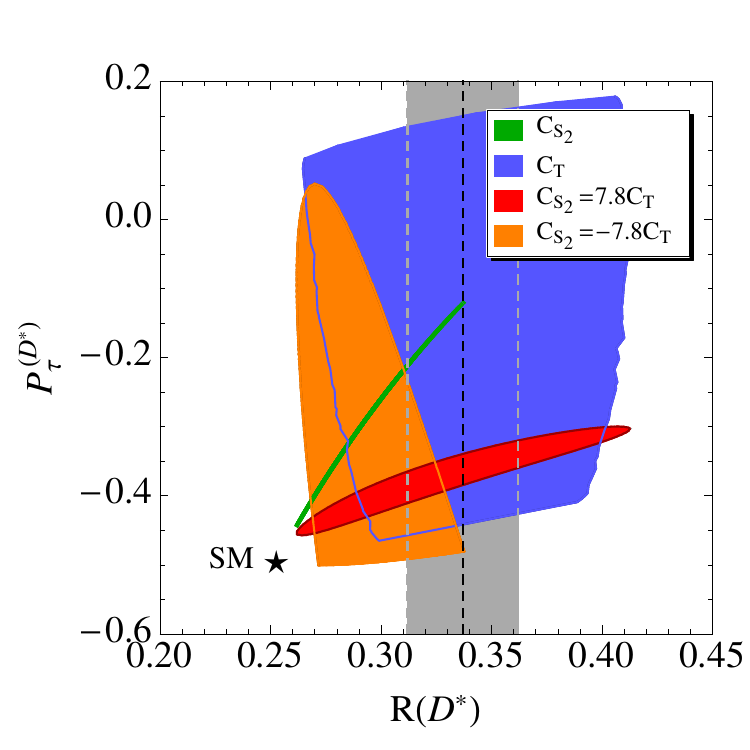} \hspace{2mm}
      \includegraphics[height=0.29\textwidth]{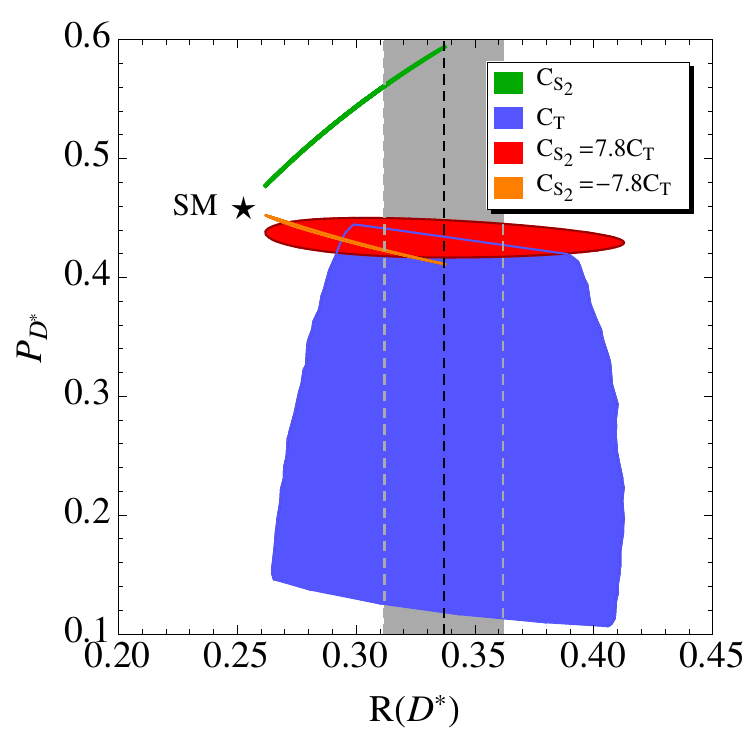}
      \\
      \vspace{2mm}
      \includegraphics[height=0.31\textwidth]{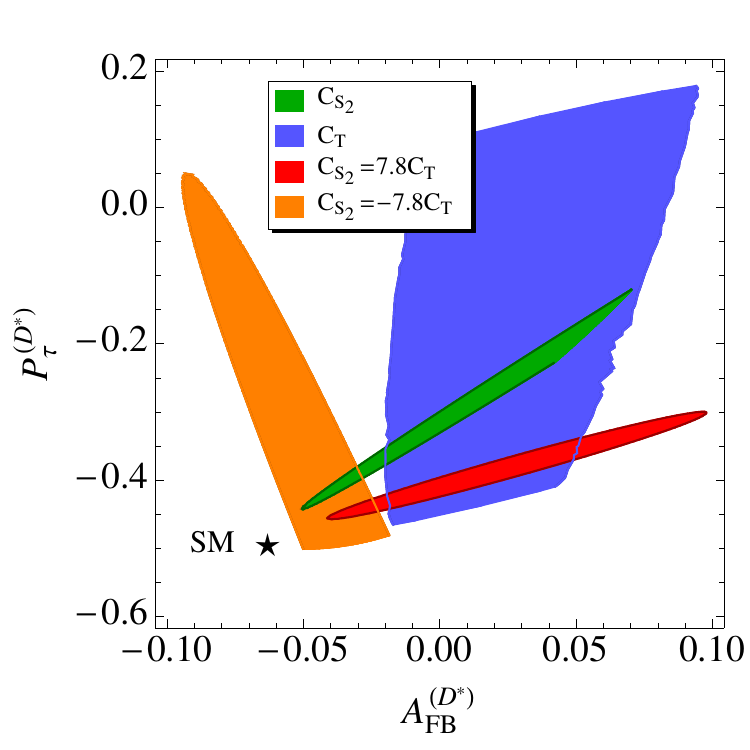} \hspace{2mm}
      \includegraphics[height=0.29\textwidth]{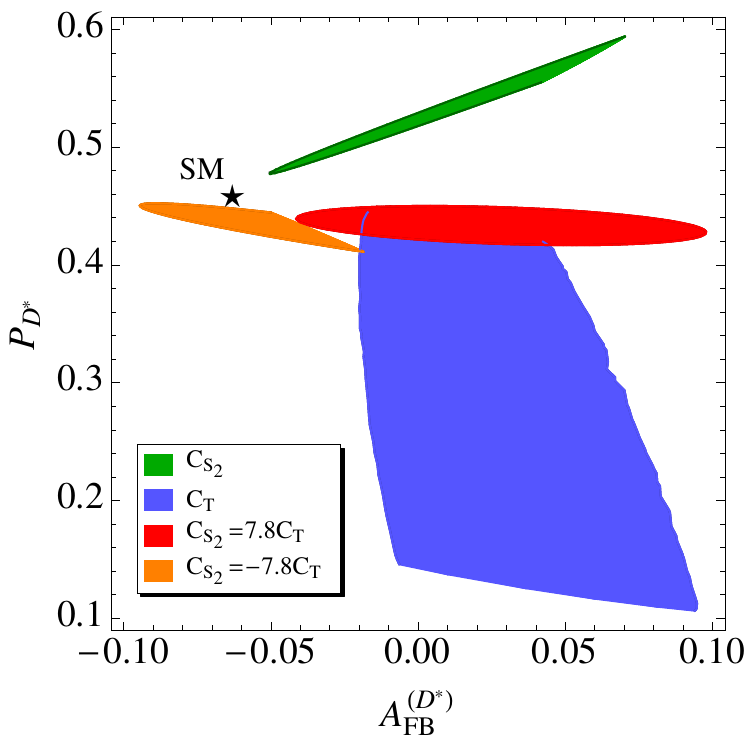} \hspace{2mm}
      \includegraphics[height=0.29\textwidth]{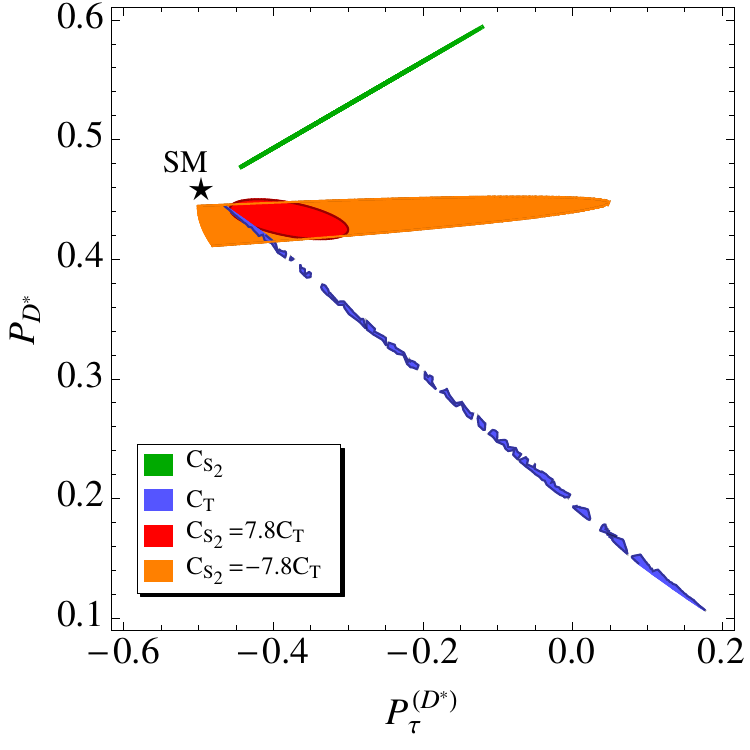}
      \caption{\footnotesize The correlations between various observables ($R(D^{(*)})$, $\A_{\rm FB}$, $P_\tau$ and $P_\Dst$) for four different NP scenarios assuming $l=\tau$: the generic scalar (green) and tensor (blue) contributions to the $C_{S_2}^\tau$ and $C_T^\tau$ Wilson coefficients respectively; only $R_2^{2/3}$ (red) and $S_1^{1/3}$ (orange) leptoquark contribution - the specific cases giving $C_{S_2}^\tau(\mu_b)=\pm7.8 C_T^\tau(\mu_b)$. The correlations were obtained by applying the constraints on the NP couplings from the $\chi^2$ fit of $R(D)$ and $R(\Dst)$ at $3\sigma$ level. The star corresponds to the SM prediction. The current experimental measurements of $R(D^{(*)})$ within $\pm1\sigma$ range are shown in gray.}
      \label{fig:correlations}
\end{figure}

In Fig. \ref{fig:correlations} we present the correlations between various observables for four different scenarios assuming $l=\tau$ \footnote{Note that the contribution to $C_{V_1}^l$ of the $U_1^{2/3}$ leptoquark, which effective couplings $h_{1L}^{2l}h_{1L}^{33*}$ remain unconstrained by $\B(\Bbar \to X_s\nu\nubar)$, gives the same asymmetry and polarizations as the SM.}:
\begin{enumerate}
   \item the generic NP scalar contribution to $C_{S_2}^\tau$ (green);
   \item the generic NP tensor contribution to $C_T^\tau$ (blue);
   \item the $R_2^{2/3}$ leptoquark contribution to $C_{S_2}^\tau$ and $C_T^\tau$ giving $C_{S_2}^\tau= 7.8 \, C_T^\tau$ (red);
   \item the $S_1^{1/3}$ leptoquark contribution to $C_{S_2}^\tau$ and $C_T^\tau$ giving $C_{S_2}^\tau=-7.8 \, C_T^\tau$ (orange). 
\end{enumerate}
The correlations are obtained by applying the constraints on the NP couplings from the $\chi^2$ fit of $R(D)$ and $R(\Dst)$ at $3\sigma$ level employing the central values of the HQET form factor parameters. The star corresponds to the SM prediction. The current experimental measurements of $R(D^{(*)})$ within $\pm1\sigma$ interval are shown in gray.

One can easily rewrite Eqs.~\eqref{eq:Ptau} and \eqref{eq:PDst} in the following forms,
\begin{equation}
   \begin{split}
      (1-P_\tau)\Gamma &= 2\Gamma(\lambda_\tau=-1/2) \,, \\
      (1-P_\Dst)\Gamma &= \Gamma(\lambda_\Dst=1) + \Gamma(\lambda_\Dst=-1) \,.
   \end{split}
   \label{eq:corr_relations}
\end{equation}
Then, we notice that the right-hand sides of Eq. \eqref{eq:corr_relations} do not contain the scalar NP contribution (see Eqs. \eqref{eq:GammaD_tau}-\eqref{eq:GammaDst_Dst}). Therefore, in the scenario 1, the correlations between $P_\tau/P_\Dst$ and $R(D^{(*)})$ are uniquely determined.

As can be seen from Fig. \ref{fig:correlations}, for some parameter spaces, one can clearly discriminate these four scenarios or at least exclude some of them. In particular, the longitudinal $\Dst$ polarization could be very useful to discriminate the models that have the generic scalar and tensor operators, $\O_{S_2}^\tau$ and $\O_T^\tau$.

\section{Conclusions}\label{sec:conclusions}

We have studied possible New Physics explanations of the observed excess of $\Bbar \to D^{(*)}\tau\nubar$ over the SM predictions focusing on the leptoquark models. It has been turned out that the $S_1^{1/3}$ scalar leptoquark with a nonvanishing product of couplings $g_{1L}^{3l}g_{1R}^{23*}$ and $R_2^{2/3}$ with $h_{2L}^{2l}h_{2R}^{33*}$ describe the present experimental data quite well. The required magnitudes of effective couplings are $O(1)$ for the leptoquark mass of 1 TeV. The interesting feature of these scenarios is that two favourable operators, namely one of the scalar operators $\mathcal{O}_{S_2}^l$ and the tensor one $\O_T^l$, simultaneously appear and their Wilson coefficients are unambiguously related as $C_{S_2}^l=\mp 4 C_T^l$ at the leptoquark mass scale.

Apart from the above two scenarios, the $U_1^{2/3}$ vector leptoquark with nonvanishing $h_{1L}^{2l}h_{1L}^{33*}$ that generates the $V-A$ operator $\O_{V_1}^l$ is also acceptable. The other scenarios in which $\O_{V_1}^l$ is induced, $S_{1(3)}^{1/3}$ with $g_{1(3)L}^{3l}g_{1(3)L}^{23*}$ and $U_3^{2/3}$ with $h_{3L}^{2l}h_{3L}^{33*}$, are hardly consistent because the experimental constraint from $B\to X_s\nu\nubar$ is mostly incompatible with those from $\Bbar \to D^{(*)}\tau\bar\nu$. The scenarios that generate the scalar operator $\O_{S_1}^l$, $V_2^{1/3}$ with $g_{2L}^{3l}g_{2R}^{23*}$ and $U_1^{2/3}$ with $h_{1L}^{2l}h_{1R}^{33*}$, are disfavoured as in the 2HDM-II.

Theoretical uncertainties in the hadronic form factors are carefully treated in our analysis. In particular, we have compared the results of two sets of the form factors, HQET and MS. These sets have rather different uncertainties although both of them describe the experimental results of $\Bbar \to D^{(*)}\ell\nubar$ and are consistent with the heavy quark symmetry. We have shown that they give similar allowed regions in the parameter space of the leptoquark models in most cases. In some cases with small probabilities, however, the degree of exclusion highly depends on the employed form factors. This means that we must be deliberate about theoretical uncertainties in New Physics contributions as well as the SM contributions in order to exclude models of New Physics.

For further tests and discrimination of the allowed leptoquark models, we have examined correlations among the $\tau$ forward-backward asymmetries $\A_{\rm FB}$, the $\tau$ polarizations $P_\tau$ and the $\Dst$ longitudinal polarization $P_{D^*}$ in some favourable cases. We have found that $P_\Dst$ is a sensitive observable to discriminate $\O_{S_2}^l$, $\O_T^l$ and their mixture. 

Measurements of these observables in addition to more precise determination of $R(D^{(*)})$ are the key issue in order to identify the origin of the present excess of $\Bbar \to D^{(*)}\tau\nubar$. LHCb and super $B$ factories are capable of exploring New Physics in this context together with the new particle search at LHC.

\section*{Acknowledgements}

This work is supported in part by the Japan Society for the Promotion of Science Grants-in-Aid for Scientific Research Nos.~20244037, 2540027 (M.T.), No.~2402804 (A.T.) and No.~248920 (R.W.). We thank David Shih and Pouya Asadi for pointing out that the previous version is incompatible with the published version.

\appendix

\vspace{1cm}

\section{Hadronic matrix elements}\label{app:FF}

\subsection{$\bm{\Bbar\to D}$}

The SM contribution is determined by the vector current operator and the relevant matrix element is written as
\begin{equation}
   \langle D(k)|\cbar\gamma_\mu b|\Bbar(p)\rangle = \left[(p+k)_\mu-{m_B^2-m_D^2 \over q^2}q_\mu\right] F_1(q^2)+q_\mu{m_B^2-m_D^2 \over q^2}F_0(q^2) \,,
   \label{eq:Fp0_parametrization}
\end{equation}
where $F_1(0)=F_0(0)$ in order to cancel the divergence at $q^2=0$. 

Using the equation of motion,
\begin{equation}
   i\partial_\mu ( \cbar \gamma^\mu b ) = ( m_b - m_c ) \cbar b \,,
\end{equation}
one can write the scalar operator matrix element as
\begin{equation}
      \langle D(k)|\cbar b|\Bbar(p)\rangle = {1 \over m_b-m_c}q_\mu\langle D(k)|\cbar\gamma^\mu b|\Bbar(p)\rangle = {m_B^2-m_D^2 \over m_b-m_c}F_0(q^2) \,.
      \label{eq:scalar_parametrization_D}
\end{equation}
In our numerical analysis we use $m_b=(4.8\pm0.2)$~GeV and $m_c=(1.4\pm0.2)$~GeV \cite{Caprini:1997mu,Melikhov:2000yu} and treat the quark masses as a source of theoretical uncertainty.

The tensor \footnote{Pseudo tensor matrix element can be evaluated using the relation $\cbar\sigma_{\mu\nu}\gamma_5b=-{i \over 2}\epsilon_{\mu\nu\alpha\beta}\cbar\sigma^{\alpha\beta}b$. In this work we use the convention $\epsilon^{0123}=-1$.} matrix element can be parametrized as
\begin{equation}
   \langle D(k)|\cbar\sigma_{\mu\nu} b|\Bbar(p)\rangle = -i ( p_\mu k_\nu - k_\mu p_\nu ) { 2F_T(q^2) \over m_B+m_D } \,.
   \label{eq:FT_parametrization}
\end{equation}

Comparing the respective matrix elements in Eqs. \eqref{eq:Fp0_parametrization}, \eqref{eq:FT_parametrization} and \eqref{eq:hpmVA_parametrization}, \eqref{eq:hT_parametrization}, one finds the following relations between the $F_{1,\,0,\,T}$ and $h_{\pm,\,T}$ form factors, usually used in the HQET (for the HQET parametrization see Appendix \ref{app:HQET}),
\begin{equation}
   \begin{split}
      F_1(q^2) =& {1 \over 2\sqrt{m_B m_D}} \left[ ( m_B + m_D ) h_+(w(q^2)) - (m_B-m_D) h_-(w(q^2)) \right] \,, \\
      F_0(q^2) =& {1 \over 2\sqrt{m_B m_D}} \left[ { ( m_B + m_D )^2 - q^2 \over m_B + m_D } \, h_+(w(q^2)) \right. \\
                & \left. \quad\quad\quad\quad\quad - { ( m_B - m_D )^2 - q^2 \over m_B - m_D } \, h_-(w(q^2)) \right] \,, \\
      F_T(q^2) =& { m_B + m_D \over 2\sqrt{m_B m_D} } h_T(w(q^2)) \,.
   \end{split}
   \label{eq:Fp0T-hpmT_relation}
\end{equation}

\subsection{$\bm{\Bbar\to \Dst}$}

The vector and axial vector operator matrix elements can be written as
\begin{equation}
   \begin{split}
      \langle\Dst(k,\varepsilon)|\cbar\gamma_\mu b|\Bbar(p)\rangle =& -i\epsilon_{\mu\nu\rho\sigma}\varepsilon^{\nu*}p^\rho k^\sigma{2V(q^2) \over m_B+m_\Dst} \,, \\
      \langle\Dst(k,\varepsilon)|\cbar\gamma_\mu\gamma_5 b|\Bbar(p)\rangle =& \varepsilon^{\mu*}(m_B+m_\Dst)A_1(q^2) - (p+k)_\mu(\varepsilon^*q){A_2(q^2) \over m_B+m_\Dst} \\
      & -q_\mu(\varepsilon^*q){2m_\Dst \over q^2}[A_3(q^2)-A_0(q^2)] \,,
   \end{split}
   \label{eq:VA_parametrization}
\end{equation}
where
\begin{equation}
   A_3(q^2) = {m_B+m_\Dst \over 2m_\Dst}A_1(q^2)-{m_B-m_\Dst \over 2m_\Dst}A_2(q^2) \,,
\end{equation}
with $A_3(0)=A_0(0)$.

The pseudo scalar matrix element can be determined by using the equation of motion,
\begin{equation}
   i\partial_\mu ( \cbar \gamma^\mu\gamma^5 b ) = -( m_b + m_c ) \cbar \gamma^5 b \,,
\end{equation}
and is given by
\begin{equation}
   \begin{split}
      \langle\Dst(k,\varepsilon)|\cbar\gamma_5 b|\Bbar(p)\rangle =& -{1 \over m_b+m_c}q_\mu\langle\Dst(k,\varepsilon)|\cbar\gamma^\mu\gamma^5 b|\Bbar(p)\rangle \\
      =& -( \varepsilon^* q ) { 2m_\Dst \over m_b + m_c }A_0(q^2) \,.
   \end{split}
   \label{eq:scalar_parametrization_Dst}
\end{equation}

The tensor operator contribution can be parametrized as
\begin{equation}
   \begin{split}
      \langle\Dst(k,\varepsilon)|\cbar\sigma_{\mu\nu} b|\Bbar(p) & \rangle = \epsilon_{\mu\nu\rho\sigma} \biggl\{ -\varepsilon^{*\rho} (p+k)^\sigma T_1(q^2) \biggr. \\
                                                                 & + \varepsilon^{*\rho} q^\sigma { m_B^2 - m_\Dst^2 \over q^2 } [ T_1(q^2) - T_2(q^2) ] \\
                                                                 & \biggl. + 2 { (\varepsilon^* \cdot q) \over q^2 } p^\rho k^\sigma \left[ T_1(q^2) - T_2(q^2) - { q^2 \over m_B^2 - m_\Dst^2 } T_3(q^2) \right] \biggr\} \,,
   \end{split}
   \label{eq:T_parametrization}
\end{equation}
where the $T_i$ form factors, commonly used in semileptonic $B$ decays, are usually determined as
\begin{equation}
   \begin{split}
      \langle\Dst(k,\varepsilon)|\cbar\sigma_{\mu\nu}q^\nu b|\Bbar(p)\rangle =& \epsilon_{\mu\nu\rho\sigma}\varepsilon^{*\nu}p^\rho k^\sigma2T_1(q^2) \,, \\
      \langle\Dst(k,\varepsilon)|\cbar\sigma_{\mu\nu}\gamma_5 q^\nu b|\Bbar(p)\rangle =& -\left[(m_B^2-m_\Dst^2)\varepsilon^{*\mu} - (\varepsilon^* q)(p+k)_\mu\right]T_2(q^2) \\
      & - (\varepsilon^* q)\left[q_\mu-{q^2 \over m_B^2-m_\Dst^2}(p+k)_\mu\right]T_3(q^2) \,.
   \end{split}
   \label{eq:T_parametrization_2}
\end{equation}

Analogously, matching Eqs.~\eqref{eq:VA_parametrization},~\eqref{eq:T_parametrization} to Eqs.~\eqref{eq:hpmVA_parametrization},~\eqref{eq:hT_parametrization}, the form factors $V$, $A_i$ and $T_i$ can be related to $h_V$, $h_{A_i}$ and $h_{T_i}$ as follows,
\begin{equation}
   \begin{split}
      V(q^2) =& { m_B + m_\Dst \over 2\sqrt{m_B m_\Dst} } \, h_V(w(q^2)) \,, \\
      A_1(q^2) =& { ( m_B + m_\Dst )^2 - q^2 \over 2\sqrt{m_B m_\Dst} ( m_B + m_\Dst ) } \, h_{A_1}(w(q^2)) \,, \\
      A_2(q^2) =& { m_B+m_\Dst \over 2\sqrt{m_B m_\Dst} } \left[ h_{A_3}(w(q^2)) + { m_\Dst \over m_B } h_{A_2}(w(q^2)) \right] \,, \\
      A_0(q^2) =& { 1 \over 2\sqrt{m_B m_\Dst} } \left[ { ( m_B + m_\Dst )^2 - q^2 \over 2m_\Dst } \, h_{A_1}(w(q^2)) \right. \\
                & -\left. { m_B^2 - m_\Dst^2 + q^2 \over 2m_B } \, h_{A_2}(w(q^2)) - { m_B^2 - m_\Dst^2 - q^2 \over 2m_\Dst } \, h_{A_3}(w(q^2)) \right] \,,
   \end{split}
   \label{eq:VA-hVA_relation}
\end{equation}

\begin{equation}
   \begin{split}
      T_1(q^2) =& { 1 \over 2\sqrt{m_B m_\Dst} } \left[ ( m_B + m_\Dst ) h_{T_1}(w(q^2)) - ( m_B - m_\Dst ) h_{T_2}(w(q^2)) \right] \,, \\
      T_2(q^2) =& { 1 \over 2\sqrt{m_B m_\Dst} } \left[ { ( m_B + m_\Dst )^2 - q^2 \over m_B + m_\Dst } \, h_{T_1}(w(q^2)) \right. \\
                & \quad\quad\quad\quad\quad\quad \left. - { ( m_B - m_\Dst )^2 - q^2 \over m_B - m_\Dst } \, h_{T_2}(w(q^2)) \right] \,, \\
      T_3(q^2) =&  { 1 \over 2\sqrt{m_B m_\Dst} } \left[ ( m_B - m_\Dst ) h_{T_1}(w(q^2)) - ( m_B + m_\Dst ) h_{T_2}(w(q^2)) \right. \\
                & \quad\quad\quad\quad\quad\quad \left.- 2 { m_B^2 -m_\Dst^2 \over m_B } h_{T_3}(w(q^2)) \right] \,.
   \end{split}
   \label{eq:T-hT_relation}
\end{equation}

\subsection{HQET form factors}\label{app:HQET}

We define the form factors of the vector and axial vector operators as
\begin{subequations}
   \label{eq:hpmVA_parametrization}
   \begin{align}
      \langle D(\vp)|\cbar \gamma_\mu b|\Bbar(v)\rangle =& \sqrt{m_B m_D} \left[ h_+(w)(v+\vp)_\mu + h_-(w)(v-\vp)_\mu \right] \,, \\
      \nonumber \\
      \begin{split}
         \langle\Dst(\vp,\varepsilon)|\cbar \gamma_\mu b|\Bbar(v)\rangle =& i\sqrt{m_B m_\Dst} h_V(w) \epsilon_{\mu\nu\rho\sigma}\varepsilon^{*\nu} v^{\prime\,\rho} v^\sigma \,, \\
         \langle\Dst(\vp,\varepsilon)|\cbar \gamma_\mu\gamma_5 b|\Bbar(v)\rangle =& \sqrt{m_B m_\Dst} \left[ h_{A_1}(w)(w+1)\varepsilon_\mu^* \right. \\
                                                                               & -\left. (\varepsilon^*\cdot v) ( h_{A_2}(w)v_\mu + h_{A_3}(w)\vp_\mu ) \right] \,,
      \end{split}
   \end{align}
\end{subequations}
where $v=p_B/m_B$, $v^\prime=k/m_{D^{(*)}}$ and $w(q^2)=v\cdot v^\prime=(m_B^2+m_{D^{(*)}}^2-q^2)/2m_Bm_{D^{(*)}}$.

In turn, using the parametrization of Caprini {\it et al.} \cite{Caprini:1997mu}, the HQET form factors can be expressed as
\begin{subequations}
   \label{eq:hpmVA_HQET}
   \begin{align}
      \begin{split}
         h_+(w) =& { 1 \over 2( 1+r_D^2-2r_D w ) } \left[ -( 1 + r_D )^2 ( w - 1 ) V_1(w) \right. \\
                 & \quad\quad\quad\quad\quad\quad\quad\quad\quad \left. + ( 1 - r_D )^2 ( w + 1 ) S_1(w) \right] \,, \\
         h_-(w) =& { ( 1 - r_D^2 ) ( w + 1 ) \over 2( 1 + r_D^2 - 2r_D w) } \left[ S_1(w) - V_1(w) \right] \,,
      \end{split} \\
      \nonumber \\
      \begin{split}
         h_V(w) =& R_1(w) h_{A_1}(w) \,, \\
         h_{A_2}(w) =& { R_2(w)-R_3(w) \over 2\,r_\Dst } h_{A_1}(w) \,, \\
         h_{A_3}(w) =& { R_2(w)+R_3(w) \over 2 } h_{A_1}(w) \,,
      \end{split}
   \end{align}
\end{subequations}
where $r_{D^{(*)}} = m_{D^{(*)}}/m_B$. The $w$-dependencies are parametrized as \cite{Caprini:1997mu}
\begin{equation}
   \begin{split}
      V_1(w) =& V_1(1) [ 1 - 8\rho_D^2 z + (51\rho_D^2-10) z^2 - (252\rho_D^2-84) z^3 ] \,, \\
      h_{A_1}(w) =& h_{A_1}(1) [ 1 - 8\rho_\Dst^2 z + (53\rho_\Dst^2-15) z^2 - (231\rho_\Dst^2-91) z^3 ] \,, \\
      R_1(w) =& R_1(1) - 0.12(w-1) + 0.05(w-1)^2 \,, \\
      R_2(w) =& R_2(1) + 0.11(w-1) - 0.06(w-1)^2 \,, \\
      R_3(w) =& 1.22 - 0.052(w-1) + 0.026(w-1)^2 \,,
      \label{eq:HQET_parametrization}
   \end{split}
\end{equation}
where $z(w) = ( \sqrt{w+1} - \sqrt2 ) / ( \sqrt{w+1} + \sqrt2 )$. The $S_1(w)$ form factor is taken from Ref.~\cite{Tanaka:2010se},
\begin{equation}
   S_1(w) = V_1(w) [ 1 + \Delta( - 0.019 + 0.041(w-1) - 0.015(w-1)^2 ) ] \,,
   \label{eq:FF_S1}
\end{equation}
with $\Delta=1 \pm 1$.

The fitted parameters, determined by the HFAG, are \cite{Amhis:2012bh}
\begin{equation}
   \begin{split}
      \rho_D^2 =& 1.186 \pm 0.054 \,, \quad~~\rho_\Dst^2 = 1.207 \pm 0.026 \,, \\
      R_1(1) =& 1.403 \pm 0.033 \,, \quad R_2(1) = 0.854 \pm 0.020 \,.
   \end{split}
   \label{eq:HQET_fit}
\end{equation}
Although the form factor normalizations, $V_1(1)$ and $h_{A_1}(1)$, vanish in the $R(D)$ and $R(\Dst)$ ratios, for completeness we provide below the latest lattice QCD calculations from Refs.~\cite{Okamoto:2004xg} and \cite{Bailey:2010gb} respectively,
\begin{equation}
   \begin{split}
      V_1(1) =& 1.074 \pm 0.024 \,, \\
      h_{A_1}(1) =& 0.908 \pm 0.017 \,.
   \end{split}
\end{equation}

The matrix elements of the tensor operator can be expressed in the following way \cite{Tanaka:2012nw},
\begin{subequations}
   \label{eq:hT_parametrization}
   \begin{align}
      \langle D(\vp)|\cbar \sigma_{\mu\nu} b|\Bbar(v)\rangle =& -i \sqrt{m_B m_D} h_T(w) [ v_\mu \vp_\nu - v_\nu \vp_\mu ] \,, \\
      \nonumber \\
      \begin{split}
         \langle \Dst(\vp,\varepsilon)|\cbar \sigma_{\mu\nu} b|\Bbar(v)\rangle =& -\sqrt{m_B m_\Dst} \epsilon_{\mu\nu\rho\sigma} \left[ h_{T_1}(w) \varepsilon^{*\rho} (v+\vp)^\sigma + h_{T_2}(w) \varepsilon^{*\rho} (v-\vp)^\sigma \right. \\
                                                                                & \quad\quad\quad\quad\quad\quad\quad\quad \left. + h_{T_3}(w) (\varepsilon^* \cdot v) (v+\vp)^\rho (v-\vp)^\sigma \right] \,,
      \end{split}
   \end{align}
\end{subequations}

As in the case of scalar operators, the equation of motion,
\begin{equation}
   \partial_\mu ( \cbar \sigma^{\mu\nu} b ) = -( m_b + m_c ) \cbar \gamma^\nu b  - (i\partial^\nu c) b + \cbar (i\partial^\nu b) \,,
\end{equation}
gives us the following relations between the tensor and vector form factors,
\begin{subequations}
   \begin{align}
      h_T(w) =& { m_b + m_c \over m_B + m_D } \left[ h_+(w) - { 1 + r_D \over 1 - r_D } \, h_-(w) \right] \,, \\
      \nonumber \\
      \begin{split}
         h_{T_1}(w) =& { 1 \over 2 ( 1 + r_\Dst^2 - 2r_\Dst w ) } \left[ { m_b - m_c \over m_B - m_\Dst } ( 1 - r_\Dst )^2 ( w + 1 ) \, h_{A_1}(w) \right. \\
                     & \quad\quad\quad\quad\quad\quad\quad\quad\quad \left. - { m_b + m_c \over m_B + m_\Dst } ( 1 + r_\Dst )^2 ( w - 1 ) \, h_V(w) \right] \,, \\
         h_{T_2}(w) =& { ( 1 - r_\Dst^2 ) ( w + 1 ) \over 2 ( 1 + r_\Dst^2 - 2r_\Dst w ) } \left[ { m_b - m_c \over m_B - m_\Dst } \, h_{A_1}(w) - { m_b + m_c \over m_B + m_\Dst } \, h_V(w) \right] \,, \\
         h_{T_3}(w) =& -{ 1 \over 2 ( 1 + r_\Dst ) ( 1 + r_\Dst^2 - 2r_\Dst w ) } \left[2 { m_b - m_c \over m_B - m_\Dst } r_\Dst ( w + 1 ) \, h_{A_1}(w) \right. \\
                     & + { m_b - m_c \over m_B - m_\Dst } ( 1 + r_\Dst^2 - 2r_\Dst w ) ( h_{A_3}(w) - r_\Dst h_{A_2}(w) ) \\
                     & \left. - { m_b + m_c \over m_B + m_\Dst } ( 1 + r_\Dst )^2 \, h_V(w) \right] \,,
      \end{split}
   \end{align}
\end{subequations}
where the residual momenta of $O(\Lambda_{\rm QCD})$ are neglected.

\subsection{Comparison of the form factors}

Here we compare three sets of form factors, evaluated in the HQET, computed by Melikhov and Stech \cite{Melikhov:2000yu}, and Cheng {\it et al.} \cite{Cheng:2003sm}. Theoretical uncertainties are not quoted directly in Refs.~\cite{Melikhov:2000yu,Cheng:2003sm}; however from the fine agreement obtained in the cases where the checks are possible, the authors of Ref.~\cite{Melikhov:2000yu} believe that the accuracy of their predictions do not exceed 10\%. Therefore, to be conservative, we vary the values of the form factors at $q^2=0$ within $\pm10\%$ around their central values. As for the HQET form factors, all theoretical parameters are supposed to have flat distributions and are randomly varied within $\pm1\sigma$ region.

\begin{figure}[t!]\centering
   \includegraphics[width=0.4\textwidth]{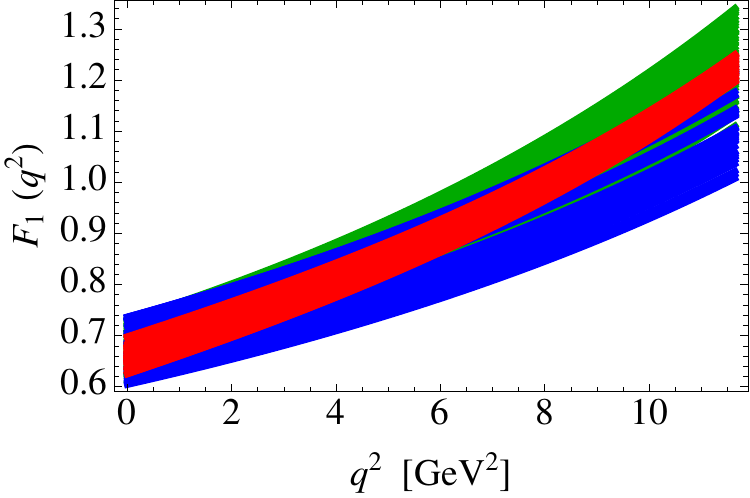} \hspace{5mm}
   \includegraphics[width=0.4\textwidth]{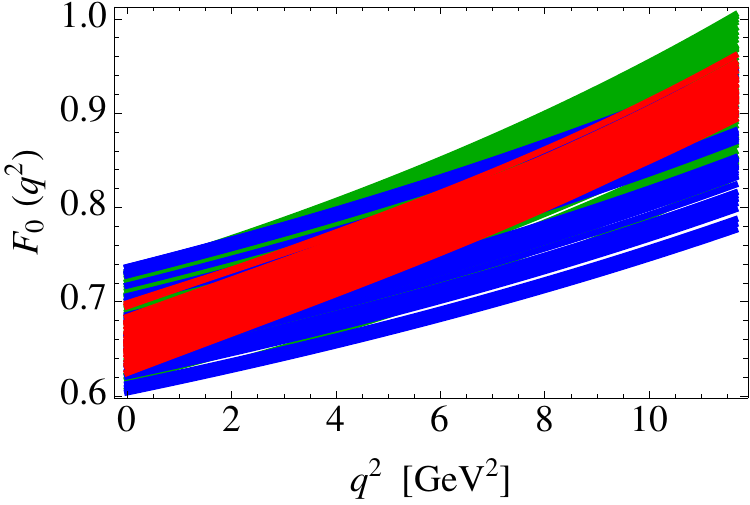}
   \vspace{5mm}
   \\
   \includegraphics[width=0.4\textwidth]{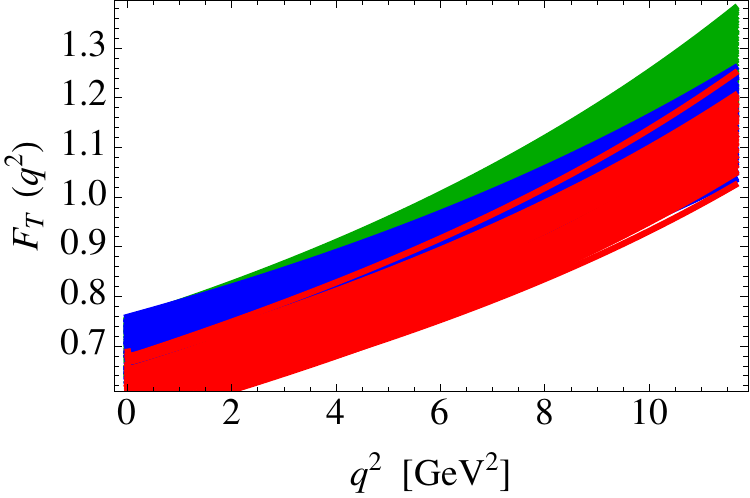}
   \caption{\footnotesize The $\Bbar\to D$ form factors evaluated in HQET (red), calculated by Melikhov and Stech \cite{Melikhov:2000yu} (blue) and by Cheng {\it et al.} \cite{Cheng:2003sm} (green). The calculation of the scalar and tensor form factors is absent in Ref. \cite{Cheng:2003sm}, therefore the equations of motion in the quark currents are used in order to express it in terms of $F_{1,0}(q^2)$.}
   \label{fig:FF_D}
\end{figure}

\begin{figure}[p!]\centering
   \includegraphics[width=0.4\textwidth]{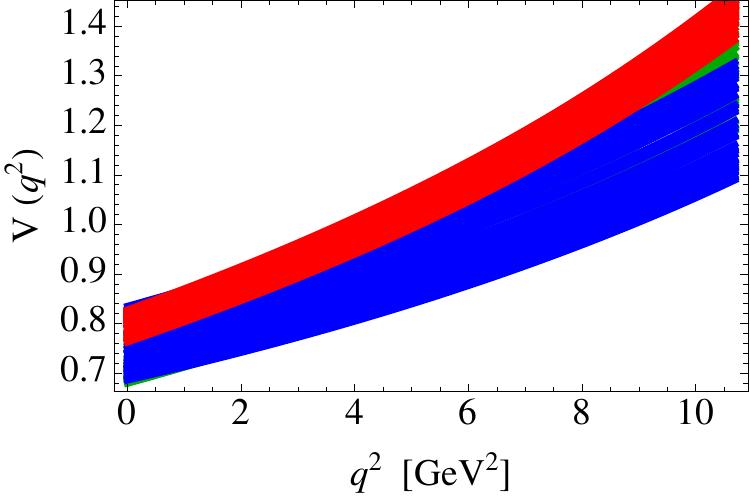} \hspace{5mm}
   \includegraphics[width=0.4\textwidth]{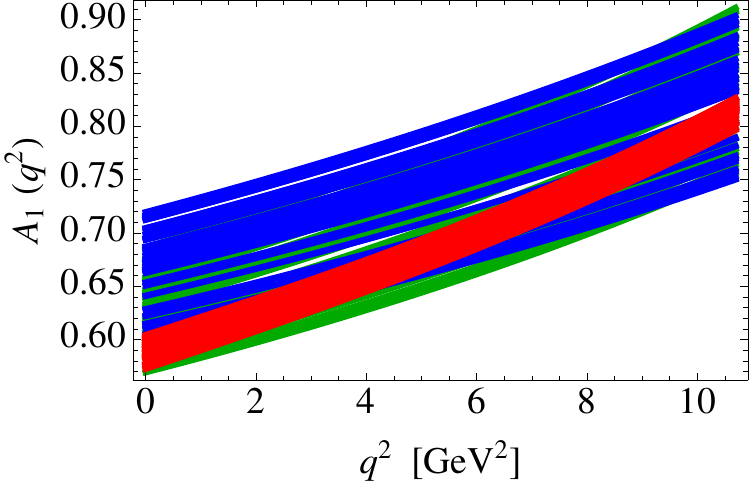}
   \vspace{5mm}
   \\
   \includegraphics[width=0.4\textwidth]{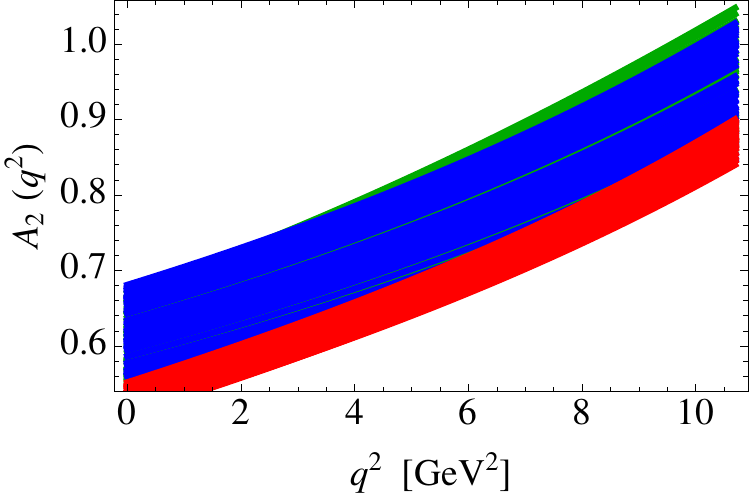} \hspace{5mm}
   \includegraphics[width=0.4\textwidth]{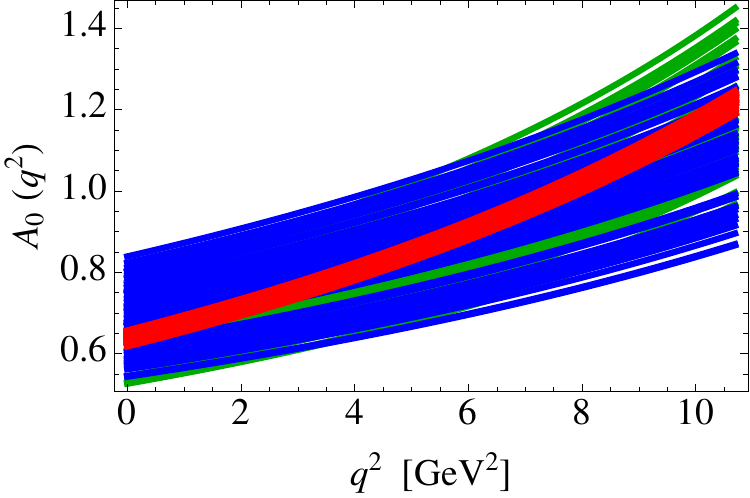}
   \vspace{5mm}
   \\
   \includegraphics[width=0.4\textwidth]{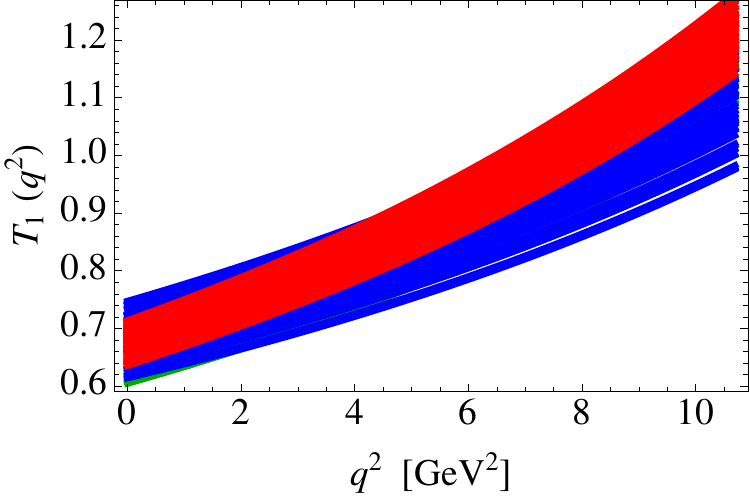} \hspace{5mm}
   \includegraphics[width=0.4\textwidth]{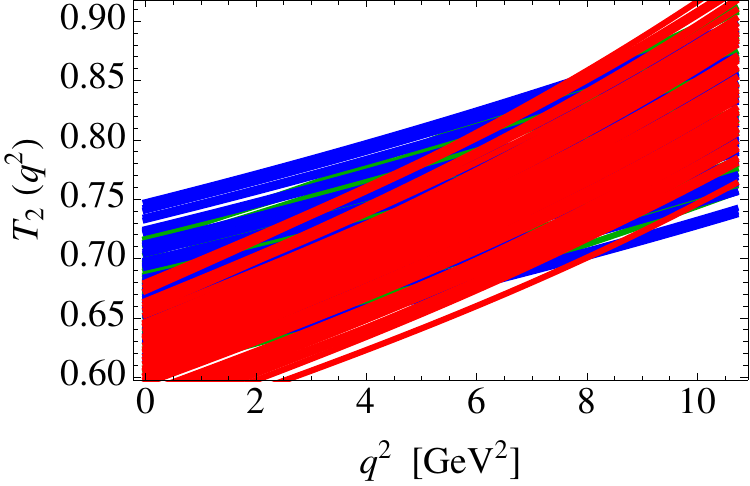}
   \vspace{5mm}
   \\
   \includegraphics[width=0.4\textwidth]{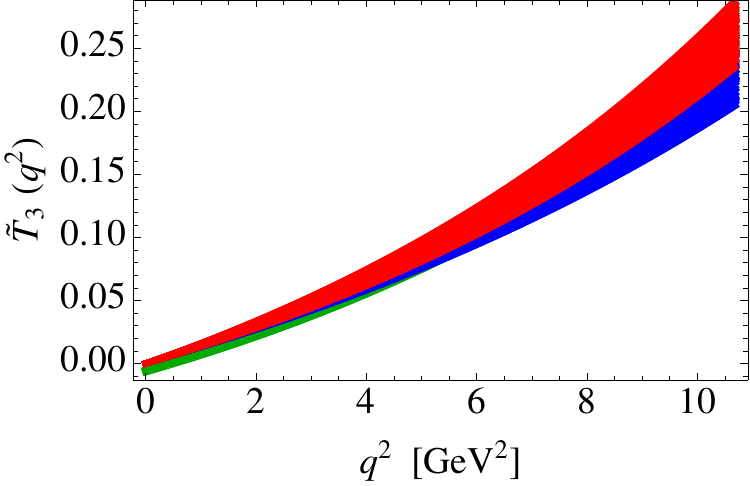}
   \caption{\footnotesize The $\Bbar\to \Dst$ form factors evaluated in HQET (red), calculated by Melikhov and Stech \cite{Melikhov:2000yu} (blue) and by Cheng {\it et al.} \cite{Cheng:2003sm} (green). The calculation of the scalar and tensor form factors is absent in Ref. \cite{Cheng:2003sm}; therefore we used the equations of motion in the quark currents in order to express $T_{1,2,3}(q^2)$ in terms of vector and axial vector form factors, $V(q^2)$ and $A_{0,1,2}(q^2)$. Here $\tilde{T}_3(q^2)$ is defined as $\tilde{T}_3(q^2)=T_3(q^2) \, q^2/(m_B^2-m_\Dst^2)$.}
   \label{fig:FF_Dst}
\end{figure}

The heavy quark limit behaviour is examined in Refs. \cite{Melikhov:2000yu,Cheng:2003sm} and the requirement of the heavy quark symmetry is satisfied. Therefore, as can be seen from Figs.~\ref{fig:FF_D} and \ref{fig:FF_Dst}, there is a reasonable agreement among these three sets.

\section{Distributions and polarizations}\label{app:distrib_and_pol}

The $q^2$ distributions for a given polarization of $\tau$ are as follows,
\begin{subequations}
   \label{eq:GammaD_tau}
   \begin{align}
      \label{eq:GammaD_tau1}
      \begin{split}
         {d\Gamma^{\lambda_\tau=1/2}(\Bbar \to D\tau\nubar_l) \over dq^2} =& {G_F^2 |V_{cb}|^2 \over 192\pi^3 m_B^3} q^2 \sqrt{\lambda_D(q^2)} \left( 1 - {m_\tau^2 \over q^2} \right)^2 \times\biggl\{ \biggr. \\
                                                                           & {1 \over 2} | \delta_{l\tau} + C_{V_1}^l + C_{V_2}^l|^2 {m_\tau^2 \over q^2} \left( H_{V,0}^{s\,2} + 3 H_{V,t}^{s\,2} \right) \\
                                                                           &+ {3 \over 2} |C_{S_1}^l + C_{S_2}^l|^2 \, H_S^{s\,2} + 8|C_T^l|^2 \, H_T^{s\,2} \\
                                                                           &+ 3\Re[ ( \delta_{l\tau} + C_{V_1}^l + C_{V_2}^l ) (C_{S_1}^{l*} + C_{S_2}^{l*} ) ] {m_\tau \over \sqrt{q^2}} \, H_S^s H_{V,t}^s \\
                                                                           &- 4\Re[ ( \delta_{l\tau} + C_{V_1}^l + C_{V_2}^l ) C_T^{l*} ] {m_\tau \over \sqrt{q^2}} \, H_T^s H_{V,0}^s \biggl.\biggr\} \,,
      \end{split} \\
      \nonumber \\
      \label{eq:GammaD_tau2}
      \begin{split}
         {d\Gamma^{\lambda_\tau=-1/2}(\Bbar \to D\tau\nubar_l) \over dq^2} =& {G_F^2 |V_{cb}|^2 \over 192\pi^3 m_B^3} q^2 \sqrt{\lambda_D(q^2)} \left( 1 - {m_\tau^2 \over q^2} \right)^2 \times\biggl\{ \biggr. \\
                                                                            & | \delta_{l\tau} + C_{V_1}^l + C_{V_2}^l|^2 \, H_{V,0}^{s\,2} + 16|C_T^l|^2 {m_\tau^2 \over q^2} \, H_T^{s\,2} \\
                                                                            &- 8\Re[ ( \delta_{l\tau} + C_{V_1}^l + C_{V_2}^l ) C_T^{l*} ] {m_\tau \over \sqrt{q^2}} \, H_T^s H_{V,0}^s \biggl.\biggr\} \,,
      \end{split}
   \end{align}
\end{subequations}

\begin{subequations}
   \label{eq:GammaDst_tau}
   \begin{align}
      \label{eq:GammaDst_tau1}
      \begin{split}
         & {d\Gamma^{\lambda_\tau=1/2}(\Bbar \to \Dst\tau\nubar_l) \over dq^2} = {G_F^2 |V_{cb}|^2 \over 192\pi^3 m_B^3} q^2 \sqrt{\lambda_\Dst(q^2)} \left( 1 - {m_\tau^2 \over q^2} \right)^2 \times\biggl\{ \biggr. \\
         & \quad\quad\quad\quad\quad {1 \over 2} ( |\delta_{l\tau} + C_{V_1}^l|^2 + |C_{V_2}^l|^2 ) {m_\tau^2 \over q^2} \left( H_{V,+}^2 + H_{V,-}^2 + H_{V,0}^2 +3 H_{V,t}^2 \right) \\
         & \quad\quad\quad\quad\quad - \Re[ (\delta_{l\tau} + C_{V_1}^l) C_{V_2}^{l*} ] {m_\tau^2 \over q^2} \left( H_{V,0}^2 + 2 H_{V,+} H_{V,-} + 3H_{V,t}^2 \right) \\      
         & \quad\quad\quad\quad\quad + {3 \over 2} |C_{S_1}^l - C_{S_2}^l|^2 \, H_S^2 + 8|C_T^l|^2 \left( H_{T,+}^2 + H_{T,-}^2 + H_{T,0}^2  \right) \\
         & \quad\quad\quad\quad\quad + 3\Re[ ( \delta_{l\tau} + C_{V_1}^l - C_{V_2}^l ) (C_{S_1}^{l*} - C_{S_2}^{l*} ) ] {m_\tau \over \sqrt{q^2}} \, H_S H_{V,t} \\
         & \quad\quad\quad\quad\quad - 4\Re[ (\delta_{l\tau} + C_{V_1}^l) C_T^{l*} ] {m_\tau \over \sqrt{q^2}} \left( H_{T,0} H_{V,0} + H_{T,+} H_{V,+} - H_{T,-} H_{V,-} \right) \\
         & \quad\quad\quad\quad\quad + 4\Re[ C_{V_2}^l C_T^{l*} ] {m_\tau \over \sqrt{q^2}} \left( H_{T,0} H_{V,0} + H_{T,+} H_{V,-} - H_{T,-} H_{V,+} \right) \biggl.\biggr\} \,,
      \end{split} \\
      \nonumber \\
      \label{eq:GammaDst_tau2}
      \begin{split}
         & {d\Gamma^{\lambda_\tau=-1/2}(\Bbar \to\Dst\tau\nubar_l) \over dq^2} = {G_F^2 |V_{cb}|^2 \over 192\pi^3 m_B^3} q^2 \sqrt{\lambda_\Dst(q^2)} \left( 1 - {m_\tau^2 \over q^2} \right)^2 \times\biggl\{ \biggr. \\
         & \quad\quad\quad\quad\quad ( |\delta_{l\tau} + C_{V_1}^l|^2 + |C_{V_2}^l|^2 ) \left( H_{V,+}^2 + H_{V,-}^2 + H_{V,0}^2 \right) \\
         & \quad\quad\quad\quad\quad - 2\Re[ (\delta_{l\tau} + C_{V_1}^l) C_{V_2}^{l*} ] \left( H_{V,0}^2 + 2 H_{V,+} H_{V,-} \right) \\      
         & \quad\quad\quad\quad\quad + 16|C_T^l|^2 {m_\tau^2 \over q^2} \left( H_{T,+}^2 + H_{T,-}^2 + H_{T,0}^2 \right) \\
         & \quad\quad\quad\quad\quad - 8\Re[ (\delta_{l\tau} + C_{V_1}^l) C_T^{l*} ] {m_\tau \over \sqrt{q^2}} \left( H_{T,0} H_{V,0} + H_{T,+} H_{V,+} - H_{T,-} H_{V,-} \right) \\
         & \quad\quad\quad\quad\quad + 8\Re[ C_{V_2}^l C_T^{l*} ] {m_\tau \over \sqrt{q^2}} \left( H_{T,0} H_{V,0} + H_{T,+} H_{V,-} - H_{T,-} H_{V,+} \right) \biggl.\biggr\} \,.
      \end{split}
   \end{align}
\end{subequations}

For the fixed polarization of $\Dst$, the distributions are given by
\begin{subequations}
   \label{eq:GammaDst_Dst}
   \begin{align}
      \label{eq:GammaDst_trDst}
      \begin{split}
         {d\Gamma^{\lambda_\Dst=\pm1}(\Bbar \to \Dst\tau\nubar_l) \over dq^2} =& {G_F^2 |V_{cb}|^2 \over 192\pi^3 m_B^3} q^2 \sqrt{\lambda_\Dst(q^2)} \left( 1 - {m_\tau^2 \over q^2} \right)^2 \times\biggl\{ \biggr. \\
                                                                               &  \left( 1 + {m_\tau^2 \over2q^2} \right) \bigl(\bigr. |\delta_{l\tau} + C_{V_1}^l|^2 H_{V,\pm}^2 + |C_{V_2}^l|^2 H_{V,\mp}^2 \\
                                                                               & \quad\quad\quad\quad\quad - 2\Re[ (\delta_{l\tau} + C_{V_1}^l) C_{V_2}^{l*}] H_{V,+} H_{V,-} \bigl.\bigr) \\
                                                                               &  + 8|C_T^l|^2 \left( 1+ {2m_\tau^2 \over q^2} \right) H_{T,\pm}^2 \\
                                                                               &  \mp 12\Re[ (\delta_{l\tau} + C_{V_1}^l) C_T^{l*} ] {m_\tau \over \sqrt{q^2}} H_{T,\pm} H_{V,\pm} \\
                                                                               &  \pm 12\Re[ C_{V_2}^l C_T^{l*} ] {m_\tau \over \sqrt{q^2}} H_{T,\pm} H_{V,\mp} \biggl.\biggr\} \,,
      \end{split} \\
      \nonumber \\
      \label{eq:GammaDst_longDst}
      \begin{split}
         {d\Gamma^{\lambda_\Dst=0}(\Bbar \to \Dst\tau\nubar_l) \over dq^2} =& {G_F^2 |V_{cb}|^2 \over 192\pi^3 m_B^3} q^2 \sqrt{\lambda_\Dst(q^2)} \left( 1 - {m_\tau^2 \over q^2} \right)^2 \times\biggl\{ \biggr. \\
                                                                            &  |\delta_{l\tau} + C_{V_1}^l - C_{V_2}^l|^2 \left[ \left( 1 + {m_\tau^2 \over2q^2} \right) H_{V,0}^2 + {3 \over 2}{m_\tau^2 \over q^2} \, H_{V,t}^2 \right] \\
                                                                            &  + {3 \over 2} |C_{S_1}^l - C_{S_2}^l|^2 \, H_S^2 + 8|C_T^l|^2 \left( 1+ {2m_\tau^2 \over q^2} \right) H_{T,0}^2 \\
                                                                            &  + 3\Re[ ( \delta_{l\tau} + C_{V_1}^l - C_{V_2}^l ) (C_{S_1}^{l*} - C_{S_2}^{l*} ) ] {m_\tau \over \sqrt{q^2}} \, H_S H_{V,t} \\
                                                                            &  - 12\Re[ ( \delta_{l\tau} + C_{V_1}^l - C_{V_2}^l) C_T^{l*} ] {m_\tau \over \sqrt{q^2}} H_{T,0} H_{V,0} \biggl.\biggr\} \,.
      \end{split}
   \end{align}
\end{subequations}

We note that the distributions for $\lambda_\tau=-1/2$ and $\lambda_\Dst=\pm1$ do not contain $C_{S_{1,2}}^l$ what makes them to be totally insensitive to the NP scalar operators.

\vskip1cm
Writing the angular distribution as
\begin{equation}
   { d^2\Gamma \over dq^2 d\cos\th } = a_\th(q^2) + b_\th(q^2) \cos\th + c_\th(q^2) \cos^2\th \,,
\end{equation}
the angular coefficient $b_\th$, which determines the lepton forward-backward asymmetry, is given by
\begin{subequations}
   \label{eq:btheta}
   \begin{align}
      \label{eq:btheta_D}
      \begin{split}
         b_\th^{(D)}(q^2) =& {G_F^2 |V_{cb}|^2 \over 128\pi^3 m_B^3} q^2 \sqrt{\lambda_D(q^2)} \left( 1 - {m_\tau^2 \over q^2} \right)^2 \times\biggl\{ \biggr. \\
                           & |\delta_{l\tau} + C_{V_1}^l + C_{V_2}^l|^2 { m_\tau^2 \over q^2} \, H_{V,0}^s H_{V,t}^s \\
                           & + \Re[ ( \delta_{l\tau} + C_{V_1}^l + C_{V_2}^l ) (C_{S_1}^{l*} + C_{S_2}^{l*} ) ] {m_\tau \over \sqrt{q^2}} \, H_S^s H_{V,0}^s \\
                           & - 4\Re[ ( \delta_{l\tau} + C_{V_1}^l + C_{V_2}^l ) C_T^{l*} ] {m_\tau \over \sqrt{q^2}} \, H_T^s H_{V,t}^s \\
                           & - 4\Re[ ( C_{S_1}^l + C_{S_2}^l ) C_T^{l*} ] \, H_T^s H_S^s \biggl.\biggr\} \,,
      \end{split} \\
      \nonumber \\
      \label{eq:btheta_Dst}
      \begin{split}
         b_\th^{(\Dst)}(q^2) =& {G_F^2 |V_{cb}|^2 \over 128\pi^3 m_B^3} q^2 \sqrt{\lambda_\Dst(q^2)} \left( 1 - {m_\tau^2 \over q^2} \right)^2 \times\biggl\{ \biggr. \\
                              & { 1 \over 2 } ( |\delta_{l\tau} + C_{V_1}^l|^2 - |C_{V_2}^l|^2 ) \left( H_{V,+}^2 - H_{V,-}^2 \right) + |\delta_{l\tau} + C_{V_1}^l - C_{V_2}^l|^2 { m_\tau^2 \over q^2 } \, H_{V,0} H_{V,t} \, \\
                              & + 8|C_T^l|^2 { m_\tau^2 \over q^2 } \left( H_{T,+}^2 - H_{T,-}^2 \right) \\
                              & + \Re[ ( \delta_{l\tau} + C_{V_1}^l - C_{V_2}^l ) (C_{S_1}^{l*} - C_{S_2}^{l*} ) ] {m_\tau \over \sqrt{q^2}} \, H_S H_{V,0} \\
                              & - 4\Re[ (\delta_{l\tau} + C_{V_1}^l) C_T^{l*} ] {m_\tau \over \sqrt{q^2}} \, \left( H_{T,0} H_{V,t} + H_{T,+} H_{V,+} + H_{T,-} H_{V,-} \right) \\
                              & + 4\Re[ C_{V_2}^l C_T^{l*} ] {m_\tau \over \sqrt{q^2}} \, \left( H_{T,0} H_{V,t} + H_{T,+} H_{V,-} + H_{T,-} H_{V,+} \right) \\
                              & - 4\Re[ ( C_{S_1}^l - C_{S_2}^l ) C_T^{l*} ] H_{T,0} H_S \biggl.\biggr\} \,.
      \end{split}
   \end{align}
\end{subequations}

\vspace{1cm}

\bibliographystyle{utphys.bst}
\bibliography{bibliography}

\providecommand{\href}[2]{#2}\begingroup\raggedright\begin{thebibliography}{10}

\bibitem{Lees:2012xj}
{\bf BaBar} Collaboration, J.~Lees {\em et al.}, ``{Evidence for an excess of
  $\bar{B} \to D^{(*)} \tau^-\bar{\nu}_\tau$ decays}'',
  \href{http://dx.doi.org/10.1103/PhysRevLett.109.101802}{{\em Phys.Rev.Lett.}
  {\bf 109} (2012)  101802},
\href{http://arxiv.org/abs/1205.5442}{{\tt arXiv:1205.5442 [hep-ex]}}.

\bibitem{Lees:2013uzd}
{\bf BaBar} Collaboration, J.~Lees {\em et al.}, ``{Measurement of an Excess of
  $B \to D^{(*)} \tau \nu$ Decays and Implications for Charged Higgs Bosons}'',
\href{http://arxiv.org/abs/1303.0571}{{\tt arXiv:1303.0571 [hep-ex]}}.

\bibitem{Matyja:2007kt}
{\bf Belle} Collaboration, A.~Matyja {\em et al.}, ``{Observation of $B^0 \to
  D^{*-} \tau^+ \nu_\tau$ decay at Belle}'',
  \href{http://dx.doi.org/10.1103/PhysRevLett.99.191807}{{\em Phys.Rev.Lett.}
  {\bf 99} (2007)  191807},
\href{http://arxiv.org/abs/0706.4429}{{\tt arXiv:0706.4429 [hep-ex]}}.

\bibitem{Adachi:2009qg}
{\bf Belle} Collaboration, I.~Adachi {\em et al.}, ``{Measurement of $B \to
  D^{(*)} \tau\nu$ using full reconstruction tags}'',
\href{http://arxiv.org/abs/0910.4301}{{\tt arXiv:0910.4301 [hep-ex]}}.

\bibitem{Bozek:2010xy}
{\bf Belle} Collaboration, A.~Bozek {\em et al.}, ``{Observation of $B^+ \to
  \bar{D}^{*0} \tau^+ \nu_\tau$ and Evidence for $B^+ \to \bar{D}^0
  \tau^+\nu_\tau$ at Belle}'',
  \href{http://dx.doi.org/10.1103/PhysRevD.82.072005}{{\em Phys.Rev.} {\bf D82}
  (2010)  072005},
\href{http://arxiv.org/abs/1005.2302}{{\tt arXiv:1005.2302 [hep-ex]}}.

\bibitem{Gunion:1989we}
J.~F. Gunion, H.~E. Haber, G.~L. Kane, and S.~Dawson, ``{THE HIGGS HUNTER'S
  GUIDE}'',
{\em Front.Phys.} {\bf 80} (2000)  1--448.

\bibitem{Martin:1997ns}
S.~P. Martin, ``{A Supersymmetry primer}'',
\href{http://arxiv.org/abs/hep-ph/9709356}{{\tt arXiv:hep-ph/9709356
  [hep-ph]}}.

\bibitem{Hou:1992sy}
W.~Hou, ``{Enhanced charged Higgs boson effects in $B^- \to \tau\bar\nu$,
  $\mu\bar\nu$ and $b \to \tau \bar\nu + X$}'',
\href{http://dx.doi.org/10.1103/PhysRevD.48.2342}{{\em Phys.Rev.} {\bf D48}
  (1993)  2342--2344}.

\bibitem{Tanaka:1994ay}
M.~Tanaka, ``{Charged Higgs effects on exclusive semitauonic $B$ decays}'',
  \href{http://dx.doi.org/10.1007/BF01571294}{{\em Z.Phys.} {\bf C67} (1995)
  321--326},
\href{http://arxiv.org/abs/hep-ph/9411405}{{\tt arXiv:hep-ph/9411405
  [hep-ph]}}.

\bibitem{Kamenik:2008tj}
J.~F. Kamenik and F.~Mescia, ``{$B \to D \tau \bar\nu$ Branching Ratios:
  Opportunity for Lattice QCD and Hadron Colliders}'',
  \href{http://dx.doi.org/10.1103/PhysRevD.78.014003}{{\em Phys.Rev.} {\bf D78}
  (2008)  014003},
\href{http://arxiv.org/abs/0802.3790}{{\tt arXiv:0802.3790 [hep-ph]}}.

\bibitem{Nierste:2008qe}
U.~Nierste, S.~Trine, and S.~Westhoff, ``{Charged-Higgs effects in a new $B \to
  D \tau\bar\nu$ differential decay distribution}'',
  \href{http://dx.doi.org/10.1103/PhysRevD.78.015006}{{\em Phys.Rev.} {\bf D78}
  (2008)  015006},
\href{http://arxiv.org/abs/0801.4938}{{\tt arXiv:0801.4938 [hep-ph]}}.

\bibitem{Tanaka:2010se}
M.~Tanaka and R.~Watanabe, ``{$\tau$ longitudinal polarization in $B \to D \tau
  \bar\nu$ and its role in the search for charged Higgs boson}'',
  \href{http://dx.doi.org/10.1103/PhysRevD.82.034027}{{\em Phys.Rev.} {\bf D82}
  (2010)  034027},
\href{http://arxiv.org/abs/1005.4306}{{\tt arXiv:1005.4306 [hep-ph]}}.

\bibitem{Fajfer:2012vx}
S.~Fajfer, J.~F. Kamenik, and I.~Nisandzic, ``{On the $B \to D^* \tau
  \bar\nu_\tau$ Sensitivity to New Physics}'',
  \href{http://dx.doi.org/10.1103/PhysRevD.85.094025}{{\em Phys.Rev.} {\bf D85}
  (2012)  094025},
\href{http://arxiv.org/abs/1203.2654}{{\tt arXiv:1203.2654 [hep-ph]}}.

\bibitem{Sakaki:2012ft}
Y.~Sakaki and H.~Tanaka, ``{Constraints of the Charged Scalar Effects Using the
  Forward-Backward Asymmetry on $B\to D^{(*)}\tau\bar{\nu_{\tau}}$}'',
  \href{http://dx.doi.org/10.1103/PhysRevD.87.054002}{{\em Phys.Rev.} {\bf D87}
  (2013)  054002},
\href{http://arxiv.org/abs/1205.4908}{{\tt arXiv:1205.4908 [hep-ph]}}.

\bibitem{Datta:2012qk}
A.~Datta, M.~Duraisamy, and D.~Ghosh, ``{Diagnosing New Physics in $b \to c \,
  \tau \, \nu_\tau$ decays in the light of the recent BaBar result}'',
  \href{http://dx.doi.org/10.1103/PhysRevD.86.034027}{{\em Phys.Rev.} {\bf D86}
  (2012)  034027},
\href{http://arxiv.org/abs/1206.3760}{{\tt arXiv:1206.3760 [hep-ph]}}.

\bibitem{Bailey:2012jg}
J.~A. Bailey, A.~Bazavov, C.~Bernard, C.~Bouchard, C.~DeTar, {\em et al.},
  ``{Refining new-physics searches in $B \to D \tau \nu$ decay with lattice
  QCD}'', \href{http://dx.doi.org/10.1103/PhysRevLett.109.071802}{{\em
  Phys.Rev.Lett.} {\bf 109} (2012)  071802},
\href{http://arxiv.org/abs/1206.4992}{{\tt arXiv:1206.4992 [hep-ph]}}.

\bibitem{Becirevic:2012jf}
D.~Becirevic, N.~Kosnik, and A.~Tayduganov, ``{$\bar B\to D\tau\bar \nu_\tau$
  vs. $\bar B\to D\mu\bar\nu_\mu$}'',
  \href{http://dx.doi.org/10.1016/j.physletb.2012.08.016}{{\em Phys.Lett.} {\bf
  B716} (2012)  208--213},
\href{http://arxiv.org/abs/1206.4977}{{\tt arXiv:1206.4977 [hep-ph]}}.

\bibitem{Fajfer:2012jt}
S.~Fajfer, J.~F. Kamenik, I.~Nisandzic, and J.~Zupan, ``{Implications of Lepton
  Flavor Universality Violations in $B$ Decays}'',
  \href{http://dx.doi.org/10.1103/PhysRevLett.109.161801}{{\em Phys.Rev.Lett.}
  {\bf 109} (2012)  161801},
\href{http://arxiv.org/abs/1206.1872}{{\tt arXiv:1206.1872 [hep-ph]}}.

\bibitem{Celis:2012dk}
A.~Celis, M.~Jung, X.-Q. Li, and A.~Pich, ``{Sensitivity to charged scalars in
  $B\to D^{(*)}\tau\nu_\tau$ and $B\to\tau\nu_\tau$ decays}'',
  \href{http://dx.doi.org/10.1007/JHEP01(2013)054}{{\em JHEP} {\bf 1301} (2013)
   054},
\href{http://arxiv.org/abs/1210.8443}{{\tt arXiv:1210.8443 [hep-ph]}}.

\bibitem{Ko:2012sv}
P.~Ko, Y.~Omura, and C.~Yu, ``{$B \to D^(*) \tau \nu$ and $B \to \tau \nu$ in
  chiral $U(1)^\prime$ models with flavored multi Higgs doublets}'',
  \href{http://dx.doi.org/10.1007/JHEP03(2013)151}{{\em JHEP} {\bf 1303} (2013)
   151},
\href{http://arxiv.org/abs/1212.4607}{{\tt arXiv:1212.4607 [hep-ph]}}.

\bibitem{Celis:2013jha}
A.~Celis, M.~Jung, X.-Q. Li, and A.~Pich, ``{$B\to D^{(*)}\tau\nu$ decays in
  two-Higgs-doublet models}'',
\href{http://arxiv.org/abs/1302.5992}{{\tt arXiv:1302.5992 [hep-ph]}}.

\bibitem{Duraisamy:2013pia}
M.~Duraisamy and A.~Datta, ``{The Full $B \to D^{*} \tau^{-} \bar{\nu_\tau}$
  Angular Distribution and $CP$ violating Triple Products}'',
\href{http://arxiv.org/abs/1302.7031}{{\tt arXiv:1302.7031 [hep-ph]}}.

\bibitem{Crivellin:2012ye}
A.~Crivellin, C.~Greub, and A.~Kokulu, ``{Explaining $B\to D\tau\nu$, $B\to
  D^*\tau\nu$ and $B\to \tau\nu$ in a 2HDM of type III}'',
  \href{http://dx.doi.org/10.1103/PhysRevD.86.054014}{{\em Phys.Rev.} {\bf D86}
  (2012)  054014},
\href{http://arxiv.org/abs/1206.2634}{{\tt arXiv:1206.2634 [hep-ph]}}.

\bibitem{Tanaka:2012nw}
M.~Tanaka and R.~Watanabe, ``{New physics in the weak interaction of $\bar B\to
  D^{(*)}\tau\bar\nu$}'',
  \href{http://dx.doi.org/10.1103/PhysRevD.87.034028}{{\em Phys.Rev.} {\bf D87}
  (2013)  034028},
\href{http://arxiv.org/abs/1212.1878}{{\tt arXiv:1212.1878 [hep-ph]}}.

\bibitem{Biancofiore:2013ki}
P.~Biancofiore, P.~Colangelo, and F.~De~Fazio, ``{On the anomalous enhancement
  observed in $B \to D^{(*)} \, \tau \, {\bar \nu}_\tau $ decays}'',
  \href{http://dx.doi.org/10.1103/PhysRevD.87.074010}{{\em Phys.Rev.} {\bf D87}
  (2013)  074010},
\href{http://arxiv.org/abs/1302.1042}{{\tt arXiv:1302.1042 [hep-ph]}}.

\bibitem{Dorsner:2013tla}
I.~Dorsner, S.~Fajfer, N.~Kosnik, and I.~Nisandzic, ``{Minimally flavored
  colored scalar in $\bar B \to D^{(*)}\tau \bar \nu$ and the mass matrices
  constraints}'',
\href{http://arxiv.org/abs/1306.6493}{{\tt arXiv:1306.6493 [hep-ph]}}.

\bibitem{Buchmuller:1986zs}
W.~Buchmuller, R.~Ruckl, and D.~Wyler, ``{Leptoquarks in lepton-quark
  collisions}'',
\href{http://dx.doi.org/10.1016/0370-2693(87)90637-X}{{\em Phys.Lett.} {\bf
  B191} (1987)  442--448}.

\bibitem{Caprini:1997mu}
I.~Caprini, L.~Lellouch, and M.~Neubert, ``{Dispersive bounds on the shape of
  $\Bbar\to D^{(*)}\ell\nubar$ form-factors}'',
  \href{http://dx.doi.org/10.1016/S0550-3213(98)00350-2}{{\em Nucl.Phys.} {\bf
  B530} (1998)  153--181},
\href{http://arxiv.org/abs/hep-ph/9712417}{{\tt arXiv:hep-ph/9712417
  [hep-ph]}}.

\bibitem{Melikhov:2000yu}
D.~Melikhov and B.~Stech, ``{Weak form-factors for heavy meson decays: An
  Update}'', \href{http://dx.doi.org/10.1103/PhysRevD.62.014006}{{\em
  Phys.Rev.} {\bf D62} (2000)  014006},
\href{http://arxiv.org/abs/hep-ph/0001113}{{\tt arXiv:hep-ph/0001113
  [hep-ph]}}.

\bibitem{ATLAS:2013oea}
{\bf ATLAS} Collaboration, G.~Aad {\em et al.}, ``{Search for third generation
  scalar leptoquarks in $pp$ collisions at $\sqrt{s}=7$ TeV with the ATLAS
  detector}'', \href{http://dx.doi.org/10.1007/JHEP06(2013)033}{{\em JHEP} {\bf
  1306} (2013)  033},
\href{http://arxiv.org/abs/1303.0526}{{\tt arXiv:1303.0526 [hep-ex]}}.

\bibitem{Chatrchyan:2012sv}
{\bf CMS} Collaboration, S.~Chatrchyan {\em et al.}, ``{Search for pair
  production of third-generation leptoquarks and top squarks in $pp$ collisions
  at $\sqrt{s}=7$ TeV}'', {\em Phys.Rev.Lett.} {\bf 110} (2013)  081801,
\href{http://arxiv.org/abs/1210.5629}{{\tt arXiv:1210.5629 [hep-ex]}}.

\bibitem{Buras:1998raa}
A.~J. Buras, ``{Weak Hamiltonian, CP violation and rare decays}'',
\href{http://arxiv.org/abs/hep-ph/9806471}{{\tt arXiv:hep-ph/9806471
  [hep-ph]}}.

\bibitem{Grossman:1995gt}
Y.~Grossman, Z.~Ligeti, and E.~Nardi, ``{New limit on inclusive $B\to
  X_s\nu\nubar$ decay and constraints on new physics}'',
  \href{http://dx.doi.org/10.1016/0550-3213(96)00051-X}{{\em Nucl.Phys.} {\bf
  B465} (1996)  369--398},
\href{http://arxiv.org/abs/hep-ph/9510378}{{\tt arXiv:hep-ph/9510378
  [hep-ph]}}.

\bibitem{Barate:2000rc}
{\bf ALEPH} Collaboration, R.~Barate {\em et al.}, ``{Measurements of $BR(b\to
  \tau^-\nubar_\tau X)$ and $BR(b\to \tau^-\nubar_\tau D^{*\pm} X)$ and upper
  limits on $BR(B^-\to \tau^-\nubar_\tau)$ and $BR(b\to s\nu\nubar)$}'',
  \href{http://dx.doi.org/10.1007/s100520100612}{{\em Eur.Phys.J.} {\bf C19}
  (2001)  213--227},
\href{http://arxiv.org/abs/hep-ex/0010022}{{\tt arXiv:hep-ex/0010022
  [hep-ex]}}.

\bibitem{Aubert:2007rs}
{\bf BaBar} Collaboration, B.~Aubert {\em et al.}, ``{Determination of the
  form-factors for the decay $B^0 \to D^{*-} \ell^{+} \nu_\ell$ and of the CKM
  matrix element $|V_{cb}|$}'',
  \href{http://dx.doi.org/10.1103/PhysRevD.77.032002}{{\em Phys.Rev.} {\bf D77}
  (2008)  032002},
\href{http://arxiv.org/abs/0705.4008}{{\tt arXiv:0705.4008 [hep-ex]}}.

\bibitem{Aubert:2008yv}
{\bf BaBar} Collaboration, B.~Aubert {\em et al.}, ``{Measurements of the
  Semileptonic Decays $\bar B \to D \ell\bar\nu$ and $\bar B \to D^*
  \ell\bar\nu$ Using a Global Fit to $D X \ell\bar\nu$ Final States}'',
  \href{http://dx.doi.org/10.1103/PhysRevD.79.012002}{{\em Phys.Rev.} {\bf D79}
  (2009)  012002},
\href{http://arxiv.org/abs/0809.0828}{{\tt arXiv:0809.0828 [hep-ex]}}.

\bibitem{Abe:2001yf}
{\bf Belle} Collaboration, K.~Abe {\em et al.}, ``{Measurement of $Br(\bar B^0
  \to D^+ \ell^-\bar\nu$) and determination of $|V_{cb}|$}'',
  \href{http://dx.doi.org/10.1016/S0370-2693(01)01483-6}{{\em Phys.Lett.} {\bf
  B526} (2002)  258--268},
\href{http://arxiv.org/abs/hep-ex/0111082}{{\tt arXiv:hep-ex/0111082
  [hep-ex]}}.

\bibitem{Dungel:2010uk}
{\bf Belle} Collaboration, W.~Dungel {\em et al.}, ``{Measurement of the form
  factors of the decay $B^0 \to D^{*-} \ell^+ \nu$ and determination of the CKM
  matrix element $|V_{cb}|$}'',
  \href{http://dx.doi.org/10.1103/PhysRevD.82.112007}{{\em Phys.Rev.} {\bf D82}
  (2010)  112007},
\href{http://arxiv.org/abs/1010.5620}{{\tt arXiv:1010.5620 [hep-ex]}}.

\bibitem{John:2012}
M.~John, ``{Prospect for $B\to D^*\tau\bar\nu_\tau$ at LHCb}'', 2012.
\newblock Talk given at
  {\href{http://indico.lal.in2p3.fr/getFile.py/access?contribId=6&sessionId=1&resId=0&materialId=slides&confId=1902}{Workshop
  on $B$ decay into $D^{**}$ and related issues}}.

\bibitem{Keune:2012phd}
A.~Keune, {\em {``Reconstruction of the Tau Lepton and the study of $B^0 \to
  D^{*-}\tau^+\nu_\tau$ at LHCb''}}.
\newblock PhD thesis, \'Ecole Polytechnique F\'ed\'erale de Lausanne, 2012.
\newblock
  \href{http://lphe.epfl.ch/publications/theses/these.an.pdf}{http://lphe.epfl.ch/publications/theses/these.an.pdf}.

\bibitem{Amhis:2012bh}
{\bf Heavy Flavor Averaging Group} Collaboration, Y.~Amhis {\em et al.},
  ``{Averages of $b$-Hadron, $c$-Hadron, and $\tau$4-lepton properties as of
  early 2012}'',
\href{http://arxiv.org/abs/1207.1158}{{\tt arXiv:1207.1158 [hep-ex]}}.

\bibitem{Okamoto:2004xg}
M.~Okamoto, C.~Aubin, C.~Bernard, C.~E. DeTar, M.~Di~Pierro, {\em et al.},
  ``{Semileptonic $D\to \pi/K$ and $B\to \pi/D$ decays in 2+1 flavor lattice
  QCD}'', \href{http://dx.doi.org/10.1016/j.nuclphysbps.2004.11.151}{{\em
  Nucl.Phys.Proc.Suppl.} {\bf 140} (2005)  461--463},
\href{http://arxiv.org/abs/hep-lat/0409116}{{\tt arXiv:hep-lat/0409116
  [hep-lat]}}.

\bibitem{Bailey:2010gb}
{\bf Fermilab Lattice Collaboration, MILC} Collaboration, J.~A. Bailey {\em et
  al.}, ``{$B\to D^* \ell\nubar$ at zero recoil: an update}'', {\em PoS} {\bf
  LATTICE2010} (2010)  311,
\href{http://arxiv.org/abs/1011.2166}{{\tt arXiv:1011.2166 [hep-lat]}}.

\bibitem{Cheng:2003sm}
H.-Y. Cheng, C.-K. Chua, and C.-W. Hwang, ``{Covariant light front approach for
  $s$ wave and $p$ wave mesons: Its application to decay constants and
  form-factors}'', \href{http://dx.doi.org/10.1103/PhysRevD.69.074025}{{\em
  Phys.Rev.} {\bf D69} (2004)  074025},
\href{http://arxiv.org/abs/hep-ph/0310359}{{\tt arXiv:hep-ph/0310359
  [hep-ph]}}.

\end{thebibliography}\endgroup

\end{document}